\documentclass[
aps,
prd,
twocolumn,
floatfix,
amsmath,
amssymb,
superscriptaddress,
notitlepage,
10pt,
preprintnumbers,
]{revtex4}

\newcounter{blind_id}
\usepackage{graphicx}
\usepackage{dcolumn}
\usepackage{bm}
\usepackage{url}
\usepackage{color}
\usepackage{xcolor}
\usepackage{comment}
\usepackage{mathrsfs}
\usepackage{multirow}
\usepackage{hyperref}

\newcommand{\dSigma}{\Delta\!\Sigma}
\newcommand{\wgg}{w_{\rm p}}
\newcommand{\avrg}[1]{\left\langle#1\right\rangle}
\newcommand{\sqdeg}{deg$^2$}

\newcommand{\commentout}[1]{}
\newcommand{\yes}{$\checkmark$}
\newcommand{\no}{--}

\begin{document}

\preprint{IPMU21-0078, YITP-21-126}

\title{HSC Year 1 cosmology results with the minimal bias method: HSC$\times$BOSS galaxy-galaxy weak lensing and BOSS galaxy clustering}

\author{Sunao~Sugiyama}
\email{sunao.sugiyama@ipmu.jp}
\affiliation{Kavli Institute for the Physics and Mathematics of the Universe (WPI), The University of Tokyo Institutes for Advanced Study (UTIAS), The University of Tokyo, Chiba 277-8583, Japan}
\affiliation{Department of Physics, The University of Tokyo, Bunkyo, Tokyo 113-0031, Japan}

\author{Masahiro~Takada}
\email{masahiro.takada@ipmu.jp}
\affiliation{Kavli Institute for the Physics and Mathematics of the Universe (WPI), The University of Tokyo Institutes for Advanced Study (UTIAS), The University of Tokyo, Chiba 277-8583, Japan}

\author{Hironao~Miyatake}
\affiliation{Kobayashi-Maskawa Institute for the Origin of Particles and the Universe (KMI),
Nagoya University, Nagoya, 464-8602, Japan}
\affiliation{Institute for Advanced Research, Nagoya University, Nagoya 464-8601, Japan}
\affiliation{Division of Particle and Astrophysical Science, Graduate School of Science, Nagoya University, Nagoya 464-8602, Japan}
\affiliation{Kavli Institute for the Physics and Mathematics of the Universe (WPI), The University of Tokyo Institutes for Advanced Study (UTIAS), The University of Tokyo, Chiba 277-8583, Japan}
\affiliation{Jet Propulsion Laboratory, California Institute of Technology,
Pasadena, CA 91109, USA}

\author{Takahiro~Nishimichi}
\affiliation{Center for Gravitational Physics, Yukawa Institute for Theoretical Physics, Kyoto University, Kyoto 606-8502, Japan}
\affiliation{Kavli Institute for the Physics and Mathematics of the Universe
(WPI), The University of Tokyo Institutes for Advanced Study (UTIAS),
The University of Tokyo, Chiba 277-8583, Japan}

\author{Masato~Shirasaki}
\affiliation{National Astronomical Observatory of Japan, Mitaka, Tokyo 181-8588, Japan}
\affiliation{The Institute of Statistical Mathematics,
Tachikawa, Tokyo 190-8562, Japan}

\author{Yosuke~Kobayashi}
\affiliation{Department of Astronomy/Steward Observatory, University of Arizona, 933 North Cherry Avenue, Tucson, AZ 85721-0065, USA}
\affiliation{Kavli Institute for the Physics and Mathematics of the Universe
(WPI), The University of Tokyo Institutes for Advanced Study (UTIAS),
The University of Tokyo, Chiba 277-8583, Japan}


\author{Surhud~More}
\affiliation{The Inter-University Centre for Astronomy and Astrophysics, Post bag 4, Ganeshkhind, Pune 411007, India}
\affiliation{Kavli Institute for the Physics and Mathematics of the Universe
(WPI), The University of Tokyo Institutes for Advanced Study (UTIAS),
The University of Tokyo, Chiba 277-8583, Japan}

\author{Ryuichi~Takahashi}
\affiliation{Faculty of Science and Technology, Hirosaki University, 3 Bunkyo-cho, Hirosaki, Aomori 036-8561, Japan}

\author{Ken~Osato}
\affiliation{Center for Gravitational Physics, Yukawa Institute for Theoretical Physics, Kyoto University, Kyoto 606-8502, Japan}
\affiliation{LPENS, D\'epartement de Physique, \'Ecole Normale Sup\'erieure,
Universit\'e PSL, CNRS, Sorbonne Universit\'e, Universit\'e de Paris, 75005 Paris, France}

\author{Masamune~Oguri}
\affiliation{Research Center for the Early Universe, The University of Tokyo, Bunkyo, Tokyo 113-0031, Japan}
\affiliation{Department of Physics, The University of Tokyo, Bunkyo, Tokyo 113-0031, Japan}
\affiliation{Kavli Institute for the Physics and Mathematics of the Universe (WPI), The University of Tokyo Institutes for Advanced Study (UTIAS), The University of Tokyo, Chiba 277-8583, Japan}

\author{Jean~Coupon}
\affiliation{Astronomy Department, University of
Geneva, Chemin d’Ecogia 16, CH-1290 Versoix, Switzerland}

\author{Chiaki~Hikage}
\affiliation{Kavli Institute for the Physics and Mathematics of the Universe
(WPI), The University of Tokyo Institutes for Advanced Study (UTIAS),
The University of Tokyo, Chiba 277-8583, Japan}

\author{Bau-Ching~Hsieh}
\affiliation{Institute of Astronomy and Astrophysics, Academia Sinica, Taipei 10617, Taiwan}

\author{Yotaka~Komiyama}
\affiliation{National Astronomical Observatory of Japan, Mitaka, Tokyo 181-8588, Japan}

\author{Alexie~Leauthaud}
\affiliation{Department of Astronomy and Astrophysics, University of California, 1156 High Street, Santa Cruz, CA 95064, USA}

\author{Xiangchong~Li}
\affiliation{Department of Physics, McWilliams Center for Cosmology, Carnegie Mellon University, Pittsburgh, PA 15213, USA}
\affiliation{Kavli Institute for the Physics and Mathematics of the Universe
(WPI), The University of Tokyo Institutes for Advanced Study (UTIAS),
The University of Tokyo, Chiba 277-8583, Japan}

\author{Wentao~Luo}
\affiliation{Department of Astronomy, School of Physical Sciences, University of Science and Technology of China, Hefei, Anhui 230026, China}
\affiliation{Key Laboratory for Research in Galaxies and Cosmology, School of Astronomy and Space Science, University of Science and Technology of China, Hefei, Anhui 230026, China}

\author{Robert~H.~Lupton}
\affiliation{Department of Astrophysical Sciences, Peyton Hall, Princeton University, Princeton, NJ 08540, USA}

\author{Hitoshi~Murayama}
\affiliation{Department of Physics, University of California, Berkeley, CA 94720, USA}
\affiliation{Kavli Institute for the Physics and Mathematics of the Universe (WPI), The University of Tokyo Institutes for Advanced Study (UTIAS), The University of Tokyo, Chiba 277-8583, Japan}
\affiliation{Ernest Orlando Lawrence Berkeley National Laboratory, Berkeley, CA 94720, USA}

\author{Atsushi~J.~Nishizawa}
\affiliation{Institute for Advanced Research, Nagoya University, Nagoya 464-8601, Japan}

\author{Youngsoo~Park}
\affiliation{Kavli Institute for the Physics and Mathematics of the Universe
(WPI), The University of Tokyo Institutes for Advanced Study (UTIAS),
The University of Tokyo, Chiba 277-8583, Japan}

\author{Paul~A.~Price}
\affiliation{Department of Astrophysical Sciences, Peyton Hall, Princeton University, Princeton, NJ 08540, USA}

\author{Melanie~Simet}
\affiliation{University of California Riverside, 900 University Ave, Riverside, CA 92521, USA}
\affiliation{Jet Propulsion Laboratory, California Institute of Technology,
Pasadena, CA 91109, USA}

\author{Joshua~S.~Speagle}
\altaffiliation{Banting \& Dunlap Fellow}
\affiliation{Department of Statistical Sciences, University of Toronto, Toronto, ON M5S 3G3, Canada}
\affiliation{David A. Dunlap Department of Astronomy \& Astrophysics, University of Toronto, Toronto, ON M5S 3H4, Canada}
\affiliation{Dunlap Institute for Astronomy \& Astrophysics, University of Toronto, Toronto, ON M5S 3H4, Canada}

\author{Michael~A.~Strauss}
\affiliation{Department of Astrophysical Sciences, Peyton Hall, Princeton University, Princeton, NJ 08544, USA}

\author{Masayuki~Tanaka}
\affiliation{National Astronomical Observatory of Japan, Mitaka, Tokyo 181-8588, Japan}




\date{\today}

\begin{abstract}
We present cosmological parameter constraints from a  blinded joint analysis of galaxy-galaxy weak lensing, $\Delta\!\Sigma(R)$, and projected correlation function, $w_\mathrm{p}(R)$, measured from the first-year HSC (HSC-Y1) data and SDSS spectroscopic galaxies over $0.15<z<0.7$. We use luminosity-limited samples as lens samples for $\Delta\!\Sigma$ and as large-scale structure tracers for $w_\mathrm{p}$ in three redshift bins, and use the HSC-Y1 galaxy catalog to define a secure sample of source galaxies at $z_\mathrm{ph}>0.75$ for the $\Delta\!\Sigma$ measurements, selected based on their photometric redshifts. For theoretical template, we use the ``minimal bias'' model for the cosmological clustering observables for the flat $\Lambda$CDM cosmological model.  We compare the model predictions with the measurements in each redshift bin on large scales, $R>12$ and $8~h^{-1}\mathrm{Mpc}$ for $\Delta\!\Sigma(R)$ and $w_\mathrm{p}(R)$, respectively, where the perturbation theory-inspired model is valid. As part of our model, we account for the effect of lensing magnification bias on the $\Delta\!\Sigma$ measurements. When we employ weak priors on cosmological parameters, without CMB information,  we find $S_8=0.936^{+0.092}_{-0.086}$, $\sigma_8=0.85^{+0.16}_{-0.11}$, and $\Omega_\mathrm{m}=0.283^{+0.12}_{-0.035}$ (mode and 68\% credible interval) for the flat $\Lambda$CDM model. Although the central value of $S_8$ appears to be larger than those inferred from other cosmological experiments, we find that the difference is consistent with expected differences due to sample variance, and 
our results are consistent with the other results to within the statistical uncertainties. When combined with the \textit{Planck} 2018 likelihood for the primary CMB anisotropy information (TT,TE,EE+lowE), we find $S_8=0.817^{+0.022}_{-0.021}$, $\sigma_8=0.892^{+0.051}_{-0.056}$, $\Omega_\mathrm{m}=0.246^{+0.045}_{-0.035}$ and the equation of state parameter of dark energy, $w_\mathrm{de}=-1.28^{+0.20}_{-0.19}$ for the flat $w$CDM model, which is consistent with the flat $\Lambda$CDM model to within the error bars.
\end{abstract}

\maketitle


\section{Introduction}\label{sect:intro}
Wide-area imaging and spectroscopic galaxy surveys provide us with powerful tools for constraining the energy composition of the universe, the growth of cosmic structure formation over time, and properties of the primordial density perturbations \citep{2013PhR...530...87W}. In particular, when combined with high-precision measurements of the cosmic microwave background \citep{wmap5,planck-collaboration:2015fj}, galaxy surveys allow us to explore the origin of the late-time cosmic acceleration, such as dark energy or a possible breakdown of General Relativity on cosmological distance scales. There are many existing, ongoing and upcoming galaxy surveys aimed at advancing our understanding of these fundamental questions; e.g.,  the SDSS-III/IV Baryon Acoustic Oscillation Spectroscopic Survey (BOSS/eBOSS) \citep{2016AJ....151...44D,2021PhRvD.103h3533A}, the Subaru Hyper Suprime-Cam (HSC) survey \cite{2018PASJ...70S...4A}, the Dark Energy Survey (DES)\footnote{https://www.darkenergysurvey.org}, the Kilo-Degree Survey (KiDS)\footnote{http://kids.strw.leidenuniv.nl}, the Subaru Prime Focus Spectrograph (PFS) survey \citep{2014PASJ...66R...1T}, the Dark Energy Spectroscopic Instrument (DESI) survey\footnote{https://www.desi.lbl.gov}, and then ultimately the Vera C.\ Rubin Observatory Legacy Survey of Space and Time (LSST)\footnote{https://www.lsst.org},  Euclid\footnote{https://sci.esa.int/web/euclid}, and the Nancy Grace Roman Space Telescope\footnote{https://roman.gsfc.nasa.gov}.

A major challenge in the use of galaxy surveys for precision cosmology lies in uncertainties in the relationship between matter and galaxy distributions in large-scale structure, i.e., uncertainties in the so-called galaxy bias \citep{kaiser84} \citep[also see][for a thorough review]{Desjacques18}. Since physical processes inherent in galaxy formation and evolution are still very challenging to accurately model from first principles, we must empirically model the galaxy bias and/or observationally constrain it. One promising observational approach is a joint probes cosmological analysis combining galaxy-galaxy weak lensing and galaxy clustering \citep{2005PhRvD..71d3511S,2009MNRAS.394..929C,2013MNRAS.432.1544M,cacciatoCombiningGalaxyClustering2012,hikage:2013kx,2015ApJ...806....2M,2017arXiv170609359K,2018PhRvD..98d3526A,2018MNRAS.476.4662V,2020MNRAS.492.2872W,2020JCAP...03..044N,Miyatake:2021b}. The two-point correlation function, $\xi_{\rm gg}$, is the standard tool to characterize the large-scale structure through galaxy clustering  \citep{peacock01,2004ApJ...606..702T}. For a cold dark matter-dominated universe with adiabatic, Gaussian initial conditions, the galaxy correlation function is related to the two-point correlation function of the underlying matter distribution, $\xi_{\rm mm}$, on large scales via a linear bias parameter as $\xi_{\rm gg}(r)\simeq b_1^2\xi_{\rm mm}(r)$ \citep{kaiser84}, where $b_1$ is a scale-independent coefficient, the value of which depends on galaxy properties \citep{dalal08}. Cross-correlating the positions of galaxies with the shapes of background galaxies as a function of their separations on the sky provides a measurement of the average matter distribution around the foreground (lensing) galaxies -- the so-called galaxy-galaxy weak lensing \citep{Mandelbaum:2005wv,mandelbaum06}. The galaxy-galaxy weak lensing arises from the galaxy-matter cross-correlation, $\xi_{\rm gm}$, which is given, at large scales, by $\xi_{\rm gm}(r)\simeq b_1\xi_{\rm mm}(r)$. Hence, combining $\xi_{\rm gg}$ and $\xi_{\rm gm}$ allows one to observationally constrain the galaxy bias for the foreground galaxy sample, at least on large scales. 

On the theory side, providing sufficiently accurate theoretical templates for accurately extracting cosmological information from the clustering observables remains difficult.  There are two competing goals for the analysis: -- ``robustness'' and ``precision'' \citep[also see Refs.][for a study based on similar motivation]{Nishimichi:2020tvu,Miyatake:2021b}. Achieving robust results requires us  to minimize any possible bias or shift in the estimated value(s) of cosmological parameter(s) from the true value(s). On the other hand, achieving precise results involves obtaining as small of a credible interval (error bars) in cosmological parameters as is possible from the  observables. Obviously it is not straightforward to achieve these two goals simultaneously. For example, since the galaxy clustering observables have a higher signal-to-noise ratio at smaller spatial scales, which are affected by nonlinear structure formation and galaxy physics, increasing the precision (reducing the error bars) in cosmological parameters requires use of the clustering observables down to small scales in the nonlinear regime. If the theoretical model is not sufficiently accurate on these small scales, it can easily lead to a large bias in the estimated cosmological parameters. The worst case scenario is that one could measure cosmological parameters that differ at high significance from the true ones. 

In this paper we show the results of cosmological parameter estimation from joint analysis of the galaxy-galaxy weak lensing, $\dSigma(R)$, and the projected correlation function of galaxies, $\wgg(R)$, measured from the Subaru HSC Year 1 data \citep[hereafter HSC-Y1 and see Refs.][for details]{2018PASJ...70S...4A,2018PASJ...70S...8A} and the spectroscopic LOWZ and CMASS galaxy samples from the Sloan Digital Sky Survey \citep[SDSS][]{Dawson:2012va}. For the theoretical template, we use the ``minimal'' bias model motivated by the perturbation theory of structure formation \citep{bernardeau02}; we model the galaxy-matter cross-correlation and the galaxy auto-correlation using a linear bias parameter and the nonlinear matter power spectrum. \citet{Sugiyama:2020kfr} evaluated the performance of this method by comparing the model predictions with simulated $\dSigma(R)$ and $\wgg(R)$ signals mimicking the HSC-Y1 and SDSS measurements. They showed that the model can recover the underlying cosmological parameters to within the statistical errors as long as the parameter inference is restricted to relatively large scales, $R\gtrsim 10\,h^{-1}{\rm Mpc}$, where the cross-correlation coefficient, defined as $r_{\rm cc}(r)\equiv \xi_{\rm gm}(r)/[\xi_{\rm gg}(r)\xi_{\rm mm}(r)]^{1/2}$, is close to unity for a $\Lambda$CDM-like cosmology. 

In our cosmological inference we perform a blind analysis to avoid confirmation biases affecting our results. After unblinding, we compare our results with the results from other cosmological experiments such as {\it Planck} and other weak lensing surveys. The results of this paper can be compared with the results in our companion paper, \citet{Miyatake:2021b}, which infers cosmological parameters by applying the halo model to exactly the same observables while including smaller-scale information than in this paper.  

This paper is organized as follows.  We briefly review the HSC-Y1 and SDSS data and catalogs used in this paper in Section~\ref{sec:data}, and then describe the measurements of $\dSigma$ and $\wgg$ in Section~\ref{sec:measurements}. Here we refer to \citet{Miyatake:2021b} for the details. In Section~\ref{sec:analysis} we describe our cosmological analysis method: the theoretical templates, the likelihood analysis, and the analysis setup. In Section~\ref{sec:blinding} we describe the blinding scheme for our cosmological analysis. In Section~\ref{sec:results} we show the resulting cosmological parameter constraints. Finally, we conclude in Section~\ref{sec:summary-and-conclusion}.

Throughout this paper we quote 68\% credible intervals for parameter uncertainties unless otherwise stated. 

\section{The SDSS and HSC-Y1 data: large-scale structure tracers and source galaxies}
\label{sec:data}

We use the data from the first-year Subaru Hyper Suprime-Cam survey \citep[hereafter HSC or HSC-Y1;][]{2018PASJ...70S...4A,2018PASJ...70S...8A} and the SDSS-III BOSS DR11 spectroscopic sample of galaxies \footnote{\url{https://www.sdss.org/dr11/.}} \citep{2015ApJS..219...12A,2013AJ....145...10D}. Hence we refer readers to \citet{Miyatake:2021b} for details, and here we briefly review the most essential aspects for this paper. 

This paper focuses on cosmological parameter inference from joint measurements of galaxy-galaxy weak lensing ($\dSigma$) and the projected galaxy auto-correlation function ($\wgg$). As tracers of large scale structure, we use luminosity-limited samples selected from the SDSS spectroscopic galaxy sample, after performing k-corrections. We consider three galaxy samples covering three distinct redshift ranges: ``LOWZ'' galaxies in the redshift range $z=[0.15,0.35]$ and two subsamples of ``CMASS'' galaxies, hereafter called ``CMASS1'' and ``CMASS2'', respectively. These are obtained by subdividing CMASS galaxies into two redshift bins, $z=[0.43,0.55]$ and $[0.55,0.70]$, respectively. More precisely, we select SDSS galaxies with absolute $i$-band magnitudes $M_i\le -21.5$, $-21.9$ and $-22.2$ for the LOWZ, CMASS1 and CMASS2 samples, respectively, yielding comoving number densities $\bar{n}_{\rm g}/[10^{-4}\,(h^{-1}{\rm Mpc})^{-3}] \simeq 1.8$, $0.74$ and $0.45$ for the $\Lambda$CDM model that is consistent with  the {\it Planck} 2015 ``TT,TE,EE+lowP'' constraints. These are lower than those of the full LOWZ and CMASS samples by a factor of a few. The luminosity-selected samples will allow for a higher-fidelity cosmology analysis because they minimize possible redshift evolution of galaxy properties within the redshift bin \citep[e.g.][]{2015ApJ...806....1M}. 

We use the HSC-Y1 galaxy catalog to define a secure sample of source galaxies behind the lens galaxies in each SDSS galaxy sample, for galaxy-galaxy weak lensing measurements. In this paper we use the HSC-Y1 shape catalog of galaxies \citep{2018PASJ...70S..25M}, in combination with photo-$z$ information \citep{2018PASJ...70S...9T}. While multiple photo-$z$ catalogs based on different methods are available, we use the catalog based on the {\tt MLZ} method as our fiducial catalog. The depth of HSC-Y1 data permits us to securely select background galaxies behind the SDSS galaxies. We define a sample of background galaxies by imposing the following cut for each HSC galaxy:
\begin{align}
\int^7_{z_{\rm l,max}+0.05}\mathrm{d}z_{\rm s}~ P_i(z_{\rm s})\ge 0.99, 
\label{eq:def_source_galaxies}
\end{align}
where $P_i(z_{\rm s})$ is the posterior distribution of photo-$z$ for the $i$-th HSC galaxy. Here $z_{\rm s}=0.75$ is chosen for the lower bound of the integration so that the HSC source galaxies are at redshifts greater than the highest redshift of SDSS galaxies in the CMASS2 sample by more than $\Delta z=0.05$, with probability greater than 0.99. For the upper bound we set $z_{\rm s}=7$, the maximum redshift adopted in the HSC photo-$z$ catalog. We use 4,308,983 HSC galaxies over about 140~\sqdeg\ in total, corresponding to $\bar{n}_{\rm s}\simeq 8.74\,{\rm arcmin}^{-2}$ for the net number density or $\bar{n}_{\rm s}\simeq 7.95\,{\rm arcmin}^{-2}$ for the weighted number density \citep[see Ref.][for the definition]{2013MNRAS.434.2121C}. The unweighted mean redshift of the sample is $\langle z_{\rm s}\rangle\simeq 1.34$.

In this paper we adopt a single population of source galaxies for galaxy-galaxy weak lensing measurements following the method in \citet{Oguri:2010vi}; in other words, we do not select source galaxies separately for each lens sample. Comparing the relative amplitudes of $\dSigma$ for the spectroscopic lens samples at different redshifts, while using the same source sample, allows for a  calibration of the average redshift of the HSC source galaxies, i.e. a self-calibration of the photo-$z$ errors. In addition, it allows for a self-calibration of the multiplicative shear biases that may remain in the HSC shape catalog, because they cause redshift-independent shifts in the $\dSigma$ amplitudes for all the lens samples. Moreover, the use of a single source population also allows for relatively straightforward treatment of the magnification bias effect on the galaxy-galaxy weak lensing, as we will discuss later.

\section{Measurements}
\label{sec:measurements}

\begin{figure*}
\begin{center}
    \includegraphics[width=2\columnwidth]{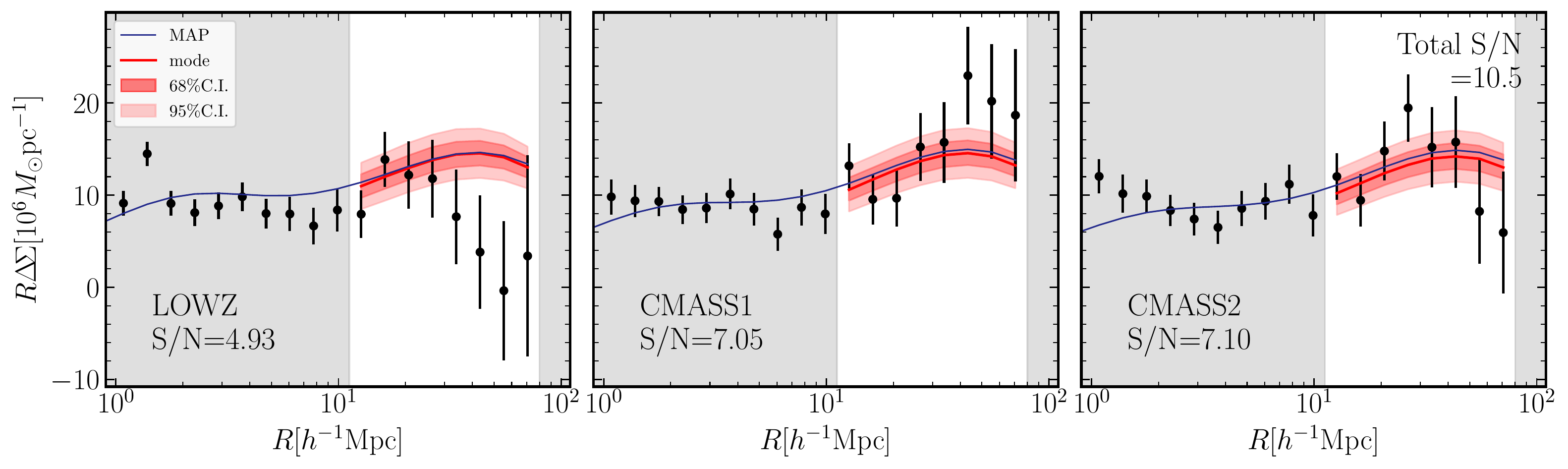}
\end{center}
\caption{The galaxy-galaxy lensing signal, 
$\Delta\!\Sigma(R)$, measured by combining the SDSS spectroscopic galaxies and the HSC photometric galaxies for lens and source galaxy samples, respectively. For illustrative purpose, we show $R\times \dSigma(R)$ so that the dynamic range of the $y$-axis is narrower. In this paper we consider three {\it luminosity-limited} lens samples, LOWZ, CMASS1 and CMASS2, in the redshift ranges $z=[0.15,0.35]$, $[0.43,0.55]$ and $[0.55,0.70]$, respectively (see text for details). The un-shaded region shows the range of separations used for cosmological parameter estimation in this paper: $R=[12,80]h^{-1}\mathrm{Mpc}$. The label in the lower left corner of each panel gives the cumulative signal-to-noise (S/N) ratio over the fitting range for each galaxy sample. The total S/N given in the upper right corner of the right panel accounts for the cross-covariances between the $\dSigma$ signals for the different lens samples. The red line in each panel indicates the mode of the posterior distributions of the model predictions in each separation bin, which are obtained from Bayesian inference applied to the $\dSigma$ and $\wgg$ data vectors for the three galaxy samples assuming a flat $\Lambda$CDM model. The dark and light red-colored regions are the 68\% and 95\% credible intervals of the posterior distribution in each bin. The posterior distributions include marginalization over other parameters. The blue line displays the model prediction at {\it maximum a posteriori} (MAP) in the Bayesian inference.
}
\label{fig:lensing-signal}
\end{figure*}
\begin{figure*}
\begin{center}
    \includegraphics[width=2\columnwidth]{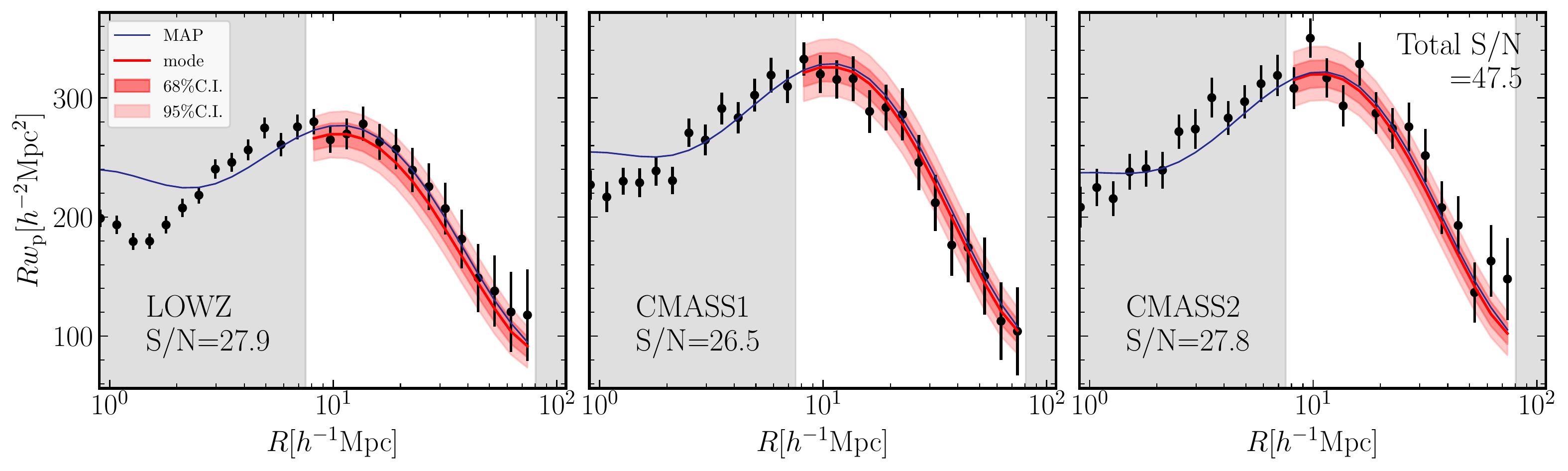}
\end{center}
\caption{Similar to the previous figure, but this figure shows the projected auto-correlation functions, $R\times \wgg(R)$, measured using the same LOWZ, CMASS1 and CMASS2 lens galaxy samples as in the previous figure. The un-shaded region denotes the range of separations used for cosmological parameter estimation: $R=[8,80]h^{-1}\mathrm{Mpc}$.}
\label{fig:clustering-signal}
\end{figure*}

The details of the measurement methods for the clustering observables, $\dSigma(R)$ and $\wgg(R)$, can be found in \citet{Miyatake:2021b}.

For the $\dSigma(R)$ measurements, we measure the average shapes of background HSC galaxies around each SDSS lens galaxy using the overlapping regions between the HSC-Y1 and SDSS survey footprints, which cover about 140~\sqdeg. The galaxy-galaxy weak lensing measures the average excess surface mass density profile around the SDSS galaxies in each redshift bin. The measurement includes a correction for the lensing efficiency, $\Sigma_{\rm cr}(z_{\rm l},z_{\rm s})$, and an estimate of the projected separation $R$ from the angular separation for each lens-source pair, which requires us to assume a reference cosmology that generally differs from the underlying true cosmology. For a flat $w\Lambda$CDM cosmology, the relevant cosmological parameters for the conversion are $\Omega_{\rm m}$ and the dark energy equation of state parameter $w_{\rm de}$. We use the method in Ref.~\cite{2013ApJ...777L..26M} to account for the $\Omega_{\rm m}$ and $w_{\rm de}$ dependences of the $\dSigma$ measurement. Note that, once the $\Sigma_{\rm cr}$ factor is included, the observable $\dSigma(R)$ depends only on the clustering properties of the SDSS lens galaxies, and does not depend on the redshifts of HSC source galaxies. 

For the $\wgg(R)$ measurements, we use the Landy \& Szalay estimator \citep{Landy:1993} to estimate the three-dimensional clustering correlation function of $\xi_{\rm gg}(R,\Pi)$ for each SDSS galaxy sample using the entire SDSS footprint covering about 8,000~\sqdeg, and then project $\xi_{\rm gg}(R,\Pi)$ over $\Pi=[-\pi_{\rm max},\pi_{\rm max}]$ to obtain the projected auto-correlation function $\wgg(R)$. We adopt $\pi_{\rm max}=100\,h^{-1}{\rm Mpc}$ as our default choice. The projected correlation function is not very sensitive to the redshift-space distortion (RSD) due to peculiar motions of galaxies, but the RSD effect is not completely negligible either \citep{vandenBosch:2012nq,Miyatake:2020uhg}. We use the method in Ref.~\cite{vandenBosch:2012nq} to include the RSD effect on $\wgg(R)$ and its cosmological dependence \citep[see also][]{Sugiyama:2020kfr}.

Figs.~\ref{fig:lensing-signal} and \ref{fig:clustering-signal} show the $\dSigma(R)$ and $\wgg(R)$ signals measured for the LOWZ, CMASS1 and CMASS2 samples. In this paper we use the signals over the range of $R$ indicated by the un-shaded region for our cosmology analysis: we use the measurements over $12<R/[h^{-1}{\rm Mpc}]<80$ (8 bins) for $\dSigma(R)$ and $8<R/[h^{-1}{\rm Mpc}]<80$ (14 bins) for $\wgg(R)$, respectively. Note that we do not include information from the baryon acoustic oscillations in $\wgg(R)$ at $R\sim 100\,h^{-1}{\rm Mpc}$ for the cosmological parameter inference. The red-colored regions in each panel indicate the 68\% and 95\% intervals of the posterior distributions of the model predictions in each separation bin, which we obtain from the cosmological parameter inference using the measurements of  $\dSigma$ and $\wgg$ for all three samples assuming a flat $\Lambda$CDM model. The blue line indicates the model prediction for the maximum a posteriori (MAP) in the parameter estimation. It is clear that the model prediction can fairly well reproduce all of the measurements simultaneously. However, one might notice a sizable discrepancy in $\dSigma$ between the MAP model prediction and the measurement, for the LOWZ sample. The measured signals at $R\gtrsim 30\,h^{-1}{\rm Mpc}$ are systematically lower than the model prediction by more than $1\sigma$ errors, although the data points are highly correlated with each other at these scales. We will below discuss in detail how the cosmological constraints are changed if the LOWZ sample is removed from the data vector in the parameter inference.

To carry out the cosmological parameter inference, we must model the covariance matrices that describe statistical uncertainties in the $\dSigma(R)$ and $\wgg(R)$ measurements. We use the jackknife method to estimate the covariance for $\wgg$. On the other hand, we use mock catalogs of SDSS lens galaxies and HSC source galaxies to model the covariance for $\dSigma$, as described in Appendix~B of \citet{Miyatake:2021b}. The covariance matrix for the $\dSigma$ data vector includes cross-correlations between the $\dSigma$ signals for different lens sample, which arise from the shape noise from the same source galaxies and the cosmic shear on the same source galaxies due to the shared foreground large-scale structure. In this and companion papers \cite{Miyatake:2021b}, we also include a contribution of the magnification bias to the covariance matrix of $\dSigma$, as derived in Appendix~\ref{sec:magbias}. We neglect the cross-covariance between $\dSigma$ and $\wgg$, because the overlapping area ($\sim $140~\sqdeg) between the HSC-Y1 and SDSS survey footprints is very small compared to the SDSS survey area ($\sim 8,000$~\sqdeg).

In the legend of Figs.~\ref{fig:lensing-signal} and \ref{fig:clustering-signal}, we give the total signal-to-noise ratio (S/N) that is obtained by integrating the S/N at each separation bin over the fitting range, $R=[12,80]$ or $[8,80]\,h^{-1}{\rm Mpc}$ for $\dSigma$ and $\wgg$, respectively, while accounting for the cross covariances between the different $R$ bins. For the total S/N for $\dSigma$, we further take into account the cross covariances between the different $\dSigma(R)$ signals for the different lens samples. The S/N values for $\wgg$ are higher than those for $\dSigma$ for each SDSS sample, meaning that our cosmological constraints are dominated by $\wgg(R)$. However, combining it with galaxy-galaxy weak lensing, $\dSigma(R)$, is critical to constrain cosmological parameters, because the combination helps break degeneracies between the galaxy bias and cosmological parameters that determine the underlying matter clustering. The total S/N of the combined (all three samples) $\dSigma$ data vector is 10.5, while the total S/N of the combined  $\wgg$ data vector is 47.5.

\section{Analysis method}
\label{sec:analysis}

In this section, we describe our cosmological parameter inference methods using the measured $\dSigma(R)$ and $\wgg(R)$ signals described in the preceding section. 

\subsection{Theoretical model}
\label{sec:model}

In this paper we adopt the ``minimal bias'' model to interpret the measured $\dSigma(R)$ and $\wgg(R)$ signals \citep[see][for details]{Sugiyama:2020kfr}. This model is the simplest one because it models $\dSigma(R)$ and $\wgg(R)$ using a linear bias parameter and the nonlinear matter correlation function. \citet{Sugiyama:2020kfr} demonstrated and validated this method by comparing theoretical templates with mock $\dSigma$ and $\wgg$ signals that mimic those of the SDSS galaxies used in this paper, showing that this method can recover the underlying cosmological parameters to within the statistical errors for the $\Lambda$CDM cosmological model that is consistent with the {\it Planck} data\citep{planck-collaboration:2015fj}, {\it as long as} the analysis is restricted to clustering information on large scales above $R_{\rm cut}=12$ and $8\,h^{-1}{\rm Mpc}$ for $\dSigma$ and $\wgg$, respectively. To be more quantitative, \citet{Sugiyama:2020kfr} showed that a shift in the central value in $S_8$ from its true value is found to be within $\sim 0.5\sigma$ ($\sigma$ is the marginalized 68\% statistical error) for all the mock signals, one of which is implemented with extremely large assembly bias effect and has a factor of 1.5 larger amplitude in $\wgg$ compared to that of the fiducial mock. The results of this paper using the minimal bias model can be compared with those of our companion paper, \citet{Miyatake:2021b}, which adopts the halo model approach to model the smaller-scale clustering signals of the SDSS galaxies down to $(2,3)~h^{-1}{\rm Mpc}$ for $\dSigma$ and $\wgg$, respectively.

Using the minimal bias model we model $\dSigma(R;z_{\rm l})$ for each of the LOWZ, CMASS1 and LOWZ2 samples as 
\begin{align}
	\Delta\!\Sigma(R;z_{\rm l})=b_1(z_{\rm l})\bar{\rho}_{\rm m0}\int_0^\infty\!\frac{k\mathrm{d}k}{2\pi}~ 
	P^{\rm NL}_{\rm mm}(k; z_{\rm l})J_2(kR),\label{eq:dsigma}
\end{align}
where $J_{2}(x)$ is the 2nd-order Bessel function, $\bar{\rho}_{\rm m0}$ is the mean matter density today, $b_1(z_{\rm l})$ is the linear galaxy bias parameter for each SDSS galaxy sample at the redshift $z_{\rm l}$, and $P_{\rm mm}^{\rm NL}$ is the nonlinear matter power spectrum at $z_{\rm l}$. In this paper, we use the {\tt halofit} fitting formula \citep{Takahashi12, Smith03} to model $P^{\rm NL}_{\rm mm}$ for an assumed cosmological model.

Throughout this paper we model the clustering observables at a representative redshift of each SDSS lens sample, defined by the mean redshift of the lens galaxies within each redshift bin. In other words, we do not include redshift evolution of the clustering observables within the redshift range. We employ $z_{\rm l}=0.26, 0.51$ and $0.63$ for the representative redshifts of the LOWZ, CMASS1 and CMASS2 samples, respectively. We checked that the difference between the signal evaluated at the representative redshift and the signal averaged over the redshift bin is below 4\% of the square root of the diagonal element of the covariance matrix in each $R$ bin, and hence this treatment should not cause any significant bias in cosmological parameters. In the following we omit the argument $z_{\rm l}$, e.g., in  $\dSigma(R;z_{\rm l})$, for notational simplicity.

The large-scale structure between us and the lens galaxies  distorts the shapes of the background source galaxies and modulates their number density. The same large-scale structure also causes number density fluctuations in the lens sample due to lensing magnification; these correlate with the source galaxy shape distortions  \citep{2020A&A...638A..96U,2021arXiv210105261V}. This effect adds a contamination term to the standard galaxy-galaxy weak lensing, expressed as
\begin{align}
\dSigma^{\rm obs}(R)=\dSigma(R)+\dSigma^{\rm mag}(R). \label{eq:add-magterm}
\end{align}
The first term on the r.h.s. is the standard galaxy-galaxy weak lensing contribution (Eq.~\ref{eq:dsigma}) and the second term is the contamination due to magnification bias, which is expressed in terms of the nonlinear matter power spectrum as
\begin{align}
\dSigma^{\rm mag}(R)&\simeq 2\left[\alpha_{\rm mag}(z_{\rm l})-1\right]
\frac{3}{2}H_0\Omega_{\rm m}
\int_0^{z_{\rm l}}\frac{\mathrm{d}z~ H_0}{H(z)}\frac{(1+z)^2}{1+z_{\rm l}}\nonumber\\
&\times \int\! {\rm d}z_{\rm s}~ P_{\rm s}(z_{\rm s}) \frac{\chi^2(\chi_{\rm l}-\chi)(\chi_{\rm s}-\chi)}{\chi_{\rm l}^2(\chi_{\rm s}-\chi_{\rm l})}\nonumber\\
&\times \bar{\rho}_{\rm m0}
\int\!\frac{k\mathrm{d}k}{2\pi}P_{\rm mm}^{\rm NL}\!(k;z)J_2\left(k\frac{\chi}{\chi_{\rm l}}R\right),
\label{eq:magnification_bias}
\end{align}
where $\alpha_{\rm mag}$ is the power-law slope of the intrinsic number counts of lens galaxies around the absolute magnitude cut, $P_{\rm s}(z_{\rm s})$ is the stacked posterior distribution of photometric redshifts for source galaxies, and $\chi_{\rm l}$ and $\chi_{\rm s}$ are the comoving angular diameter distances to the lens redshift and the source redshift, respectively. Note that $\dSigma^{\rm mag}$ does not depend on the galaxy bias. We include the redshift distribution of source galaxies, $P_{\rm s}(z_{\rm s})$, but use the mean redshift of lens galaxies for simplicity. An estimate of the number counts slope, $\alpha_{\rm mag}$, for each of the absolute magnitude-limited samples of LOWZ, CMASS1 and CMASS2 is relatively straightforward, compared to the original parent samples, which have color-dependent flux cuts that make it challenging to determine the slope for the magnification bias calculation \citep{2021arXiv210105261V}.

From the measured number counts, we estimate $\alpha_{\rm mag}\simeq 2.26\pm 0.03$, $3.56\pm 0.04$ and $3.73\pm 0.04$ for the LOWZ, CMASS1 and CMASS2 sample, respectively, where the $1\sigma$ error is estimated assuming Poisson errors in the number counts in each magnitude bin around the magnitude cut. Although we have a relatively accurate estimate of $\alpha_{\rm mag}$ for each sample, we employ a conservative approach for our cosmological analysis: we treat $\alpha_{\rm mag}$ as a nuisance parameter in the cosmological analysis, employing a Gaussian prior with width $\sigma(\alpha_{\rm mag})=0.5$ and a mean value taken from the above measurement value. As we will show below, $\dSigma^{\rm mag}$ constitutes a 1\%, 7\% and 10\% contribution to the total $\dSigma^{\rm model}$ for the LOWZ, CMASS1 and CMASS2 sample, respectively, for the {\it Planck} cosmology\footnote{Exactly speaking, we should use the {\it intrinsic} redshift distribution of source galaxies for computation of the magnification bias effect (Eq.~\ref{eq:magnification_bias}), rather than the stacked posterior distribution of photometric redshifts. However, even if we use the ``reweighted'' distribution based on the COSMOS catalog, which gives an estimate of the intrinsic distribution, it changes the magnification bias only by up to 5\% fractional changes (therefore by up to $0.5\%(=0.1\times 0.05)$ contamination to the galaxy-galaxy lensing amplitude). In addition, we employ a broad prior width of $\sigma(\alpha_{\rm mag})=0.5$ as our fiducial choice, and the $\pm 1\sigma$ shift from the central value in $\alpha_{\rm mag}$ leads to $\pm 20$--$40\%$ fractional changes in Eq.~(\ref{eq:magnification_bias}). Hence the prior width in $\alpha_{\rm mag}$ absorbs the impact of inaccuracy in the model calculation of Eq.~(\ref{eq:magnification_bias}) on the cosmological parameters.}. Including the $\dSigma^{\rm mag}$ contribution in the theoretical template adds some cosmological information. Encouragingly, we will show that the broad prior on $\alpha_{\rm mag}$ causes almost no degradation in cosmological parameter constraints.

The projected auto-correlation function $\wgg(R)$ is defined in terms of the three-dimensional correlation function $\xi_{\rm gg}$ as 
\begin{align}
	w_{\rm p}(R;z_{\rm l}) =& 2f_{\rm corr}^{\rm RSD}(R;z_{\rm l})\nonumber\\
	&\times\int_{0}^{\pi_{\rm max}}\!{\rm d}\Pi~ \xi_{\rm gg}\left(\sqrt{R^2 + \Pi^2};z_{\rm l}\right),\label{eq:wp}
\end{align}
where $\pi_{\rm max}$ is the projection length along the line of sight. We employ $\pi_{\rm max}=100\,h^{-1}{\rm Mpc}$ as used in the measurement. The projected correlation function is less sensitive to the RSD effect compared to $\xi_{\rm gg}$, because the line-of-sight projection reduces the RSD effect, but the effect is not completely negligible, especially at large $R$. We account for the residual RSD effect using the Kaiser RSD factor \cite{kaiser84}: $f_{\rm corr}^{\rm RSD}(R;z_{\rm l})$ is the correction factor, which depends on redshift and on the assumed $\Omega_{\rm m}$ and $w_{\rm de}$ for a flat $w$CDM model \cite[see Eq.~(48) in Ref.][for the definition]{vandenBosch:2012nq}. Using the minimal bias model, we model the three-dimensional, real-space galaxy correlation function as 
\begin{align}
    \xi_{\rm gg}(r) =b_1^2 \int_0^\infty\!\frac{k^2\mathrm{d}k}{(2\pi^2)}~P^{\rm NL}_{\rm mm}(k)j_0(kr),\label{eq:xi}
\end{align}
where $j_0(x)$ is the zeroth-order spherical Bessel function, and $b_1$ is the same bias parameter as in Eq.~(\ref{eq:dsigma}) for each SDSS galaxy sample. While the galaxy clustering signal is also affected by magnification bias, we have checked that its contribution is at the sub-percent level compared to Eq.~(\ref{eq:wp}), and hence neglect it in our model.

Throughout this paper, we use the logarithmic center of each separation bin as a representative projected separation, $R_i=(R_{i, {\rm max}}R_{i, {\rm min}})^{1/2}$, and evaluate the model prediction at the representative projected separation. We have checked that the difference between the model signal evaluated at the representative separation and the area-averaged model signal within each separation bin is below 2\% of the square root of the diagonal element of the covariance matrix.

In this minimal bias model, the relation between the real-space correlation functions $\xi_{\rm gm}(r)/\sqrt{\xi_{\rm gg}(r)\xi_{\rm mm}(r)}=1$ always holds by construction. Hence as long as the correlation functions for real galaxies follow this relation for large separations, our model can accurately model the measured correlation functions at such separations. 

We use \texttt{Dark Emulator} \citep{Nishimichi:2020tvu} to compute the linear matter power spectrum, $P_{\rm lin}(k)$, for an input cosmological model, which is constructed from the \texttt{CLASS} \cite{class1} outputs. We then use the updated version of halofit \citep{smith:2008yu,Takahashi12} to compute the nonlinear matter power spectrum, $P_{\rm NL}(k)$, from the linear power spectrum for the cosmological model. We use the {\tt FFTLog} method \cite{Hamilton00}, implemented in \texttt{pyfftlog} \footnote{A publicly-available python package of {\tt FFTLog}, \url{https://github.com/prisae/pyfftlog}.}, to perform the Hankel transforms in Eqs.~(\ref{eq:dsigma}) and (\ref{eq:xi}). The integration for the galaxy clustering projection in Eq.~(\ref{eq:wp}) is performed with the trapezoidal rule. The model parameters are summarized in Table~\ref{tab:parameters}.

\begin{table}
\caption{Model parameters and priors used in our cosmological parameter inference. The label ${\cal U}(a,b)$ denotes a uniform (or equivalently flat) distribution with minimum $a$ and maximum $b$, while ${\cal N}(\mu, \sigma)$ denotes a normal distribution with mean $\mu$ and width $\sigma$. The parameters above ``Extended model'' are the parameters used in our baseline analysis: 5 cosmological parameters, a linear galaxy bias parameter and a magnification bias parameter for each of the LOWZ, CMASS1 and CMASS2 samples, and 2 nuisance parameters to model residual photo-$z$ and multiplicative shear errors: $13=5+3+3+2$ in total. We perform the analysis for the ``Extended model'' after unblinding; we further include the dark energy equation of state parameter, $w_{\rm de}$, and combine the information from $\dSigma$ and $\wgg$ with  external datasets (e.g. {\it Planck}) to estimate all of the cosmological parameters in the extended model, $w$CDM.}
\label{tab:parameters}
\begin{center}
\begin{tabular}{cc}  \hline\hline
Parameter & Prior \\ \hline
\multicolumn{2}{c}{{\bf Cosmological parameters}}\\
$\Omega_{\rm de}$       & ${\cal U}(0.4594, 0.9094)$\\
$\ln(10^{10}A_{\rm s})$ & ${\cal U}(1.0,5.0)$\\
$\omega_{\rm b}$        & ${\cal N}(0.02268,0.00038)$\\
$\omega_{\rm c}$        & ${\cal U}(0.0998, 0.1398)$\\ 
$n_{\rm s}$             & ${\cal N}(0.9649,3\times0.0042)$\\ \hline
\multicolumn{2}{c}{\bf galaxy bias parameters}\\
$b_1(z_i)$              & ${\cal U}(0.1,5.0)$\\ \hline
\multicolumn{2}{c}{\bf magnification bias parameters}\\
$\alpha_{\rm mag}(z_{\rm LOWZ})$   & ${\cal N}(2.259, 0.5)$ \\
$\alpha_{\rm mag}(z_{\rm CMASS1})$ & ${\cal N}(3.563, 0.5)$ \\
$\alpha_{\rm mag}(z_{\rm CMASS2})$ & ${\cal N}(3.729, 0.5)$ \\\hline
\multicolumn{2}{c}{\bf Photo-$z$/Shear errors}\\
$\Delta z_{\rm ph}$     & ${\cal N}(0.0,0.1)$  \\
$\Delta m_\gamma$       & ${\cal N}(0.0,0.01)$ \\ \hline\hline
\multicolumn{2}{c}{\bf Extended model} \\
\hline
\multicolumn{2}{c}{after unblinding} \\
$w_{\rm de}$                     & ${\cal U}(-4.0,-0.2)$ \\
\hline\hline
\end{tabular}
\end{center}
\end{table}

\subsection{Modeling residual systematic errors in the galaxy-galaxy weak lensing}
\label{sec:systematics}

\citet{Miyatake:2021b} performed various systematics tests and null tests such as the $B$-mode signal and the ``boost'' factor, and did not find any strong evidence for residual systematic effects in the $\dSigma$ measurements, reflecting the high quality of the HSC-Y1 data. In our cosmological analysis, we introduce nuisance parameters $\Delta z_{\rm ph}$ and $\Delta m_\gamma$ to model possible residual systematic errors in the photo-$z$ and multiplicative shear calibration, and adopt conservative priors on those nuisance parameters. Hence, even if we have residual unknown systematic effects in the weak lensing measurements, the nuisance parameters can at least partially absorb their impact on the cosmological constraints. We assume that the SDSS spectroscopic data has a better control of the systematic effects, and we do not consider any residual systematic errors in the $\wgg$ measurement.

\subsubsection{A residual systematic photo-$z$ error: $\Delta z_{\rm ph}$}
\label{sec:systeamtics_nuisance}
Photo-$z$ uncertainties are among the most important systematic effects in weak lensing measurements. To study the impact of photo-$z$ errors on our results, we introduce a nuisance parameter $\Delta z_{\rm ph}$ to model the possible residual uncertainty. More specifically, following the method in Ref.~\cite{Hutereretal:06} \citep[also see Ref.][for the detailed discussion on this modeling of systematics]{Miyatake:2021b}, we model the systematical error in the mean source redshift by shifting the posterior distribution of each source galaxy by the same amount, $\Delta z_{\rm ph}$, as
\begin{align}
P_{\rm s}(z_{\rm s}) \longrightarrow P_{\rm s}(z_{\rm s}+\Delta z_{\rm ph}).
\end{align}
We then use the shifted distribution to compute the averaged lensing efficiency $\avrg{\Sigma_{\rm cr}^{-1}}_{\rm ls}$ and the weight $w_{\rm ls}$ for the source-lens pairs using the actual HSC-Y1 and SDSS catalogs. We found that the lensing signal after this shift can be well approximated by the following multiplicative form: 
\begin{align}
\widehat{\dSigma}^{(i_{\rm l})}\!\!(R;\Delta z_{\rm ph})
\simeq f_{\rm ph}^{(i_{\rm l})}\!(\Delta z_{\rm ph})
\widehat{\dSigma}^{(i_{\rm l})}\!\!(R;\Delta z_{\rm ph}=0),
\end{align}
where $f_{\rm ph}^{(i_{\rm l})}(\Delta z_{\rm })$ is a multiplicative factor to model the effect of systematic photo-$z$ error and $i_{\rm l}=$``LOWZ'', ``CMASS1'' or ``CMASS2'' is the index of the lens galaxy sample \footnote{Here, we implicitly include the cosmological dependence of $\langle\Sigma_{\rm cr}^{-1}\rangle_{\rm ls}$ in the photo-$z$ correction, by applying both a photo-$z$ correction and a measurement correction, i.e., accounting for the $\Omega_{\rm m}$ and $w_{\rm de}$ dependencies of the measurement described in Section~\ref{sec:measurements}, at the same time.}. While using a single population of source galaxies, we find that the shift $\Delta z_{\rm ph}$ leads to different changes in the amplitudes of $\dSigma$ for the different lens samples (LOWZ, CMASS1, and CMASS2) depending on the lens redshift. Conversely, we can use the relative variations in the $\dSigma$ amplitudes at different lens redshifts to calibrate out $\Delta z_{\rm ph}$, simultaneously with cosmological parameter estimation, if the data has sufficiently high signal-to-noise ratio. This is a self-calibration method for photo-$z$ error that was proposed in \citet{Oguri:2010vi}. We will see how  effectively the self-calibration method works to calibrate photo-$z$ errors to higher precision than the prior, given the statistical power of the HSC-Y1 data and the scale cuts adopted in this paper.

Exactly speaking, we have to use a shift in the intrinsic redshift distribution of source galaxies, rather than the posterior distribution, to estimate the impact of residual photo-$z$ errors on $\dSigma$. As discussed in \citet{Miyatake:2021b}, even if we use the intrinsic distribution estimated by matching the source galaxies to the calibration sample of COSMOS catalog in magnitude-color space, the effect is very small, compared to the change caused by using the parameter $\Delta z_{\rm ph}$. Hence we conclude that our treatment of the residual photo-$z$ error is sufficient to capture the possible impact. 

In our method, we apply the inverse of the photo-$z$ error factor as a correction to the theoretical template for $\dSigma(R)$, rather than changing the measurement, as follows:
\begin{align}
\dSigma^{\rm model}(R)\rightarrow \frac{\dSigma^{\rm model}(R)}{f_{\rm ph}^{(i_{\rm l})}(\Delta z_{\rm ph})}.
\label{eq:dSigma_dz_ph}
\end{align}
This treatment permits us to use an unchanged data vector and its covariance matrix for the cosmological parameter inference.

\begin{table}
\caption{Differences in the averaged $\avrg{\Sigma_{\rm cl}^{-1}}$ using different photo-$z$ method relative to that for the fiducial photo-$z$ method ({\tt MLZ}). We use the same method given in Eq.~(\ref{eq:def_source_galaxies}) to select source galaxies, and compute the average $\avrg{\Sigma_{\rm cl}^{-1}}$ over all lens-source pairs in the separation $3\le R/[h^{-1}{\rm Mpc}]\le 30$, for each lens sample (LOWZ, CMASS1 and CMASS2). The number after ``$\pm$'' denotes the $1\sigma$ uncertainty in the difference, estimated from the width of the photo-$z$ posterior of each source galaxy. 
}
\label{tab:photo-z_test}
\begin{center}
\begin{tabular}{l|ccc}
\hline\hline
photo-$z$ & LOWZ & CMASS1 & CMASS2 \\ 
method & \\
\hline 
{\tt DEmP} &$-0.048 \pm 0.034$& $-0.000 \pm 0.029$& $-0.030 \pm 0.026$\\ 
{\tt Ephor AB} &$-0.046 \pm 0.035$& $0.052 \pm 0.033$& $0.129 \pm 0.055$\\ 
{\tt Franken-Z} &$-0.003 \pm 0.023$ & $0.002 \pm 0.027 $& $0.003 \pm 0.030$\\ 
{\tt Mizuki} &$-0.042\pm 0.022$ & $-0.050 \pm 0.011 $ & $-0.041 \pm 0.012$\\ 
{\tt NNPZ} &$-0.049 \pm 0.030$& $-0.018 \pm 0.043$& $0.061 \pm 0.050$\\ 
\hline
\end{tabular}
\end{center}
\end{table}
As another sanity check, we also study the impact of different photo-$z$ methods on the cosmological results. Table~\ref{tab:photo-z_test} gives differences in $\avrg{\Sigma_{\rm cr}^{-1}}$ when using photo-$z$ estimate for each source galaxy based on different photo-$z$ methods, relative to that for the fiducial photo-$z$ method. Note that we repeated the cut defined by Eq.~(\ref{eq:def_source_galaxies}) to define the source galaxy sample for each method, so the source samples are different for each photo-$z$ method. The different photo-$z$ methods give a few percent change in the $\dSigma$ amplitudes; these changes differ depending on the lens samples. We will explicitly study to what extent the cosmological results are changed by using different photo-$z$ catalogs.

\subsubsection{A residual error in multiplicative shear bias: $\Delta m_{\gamma}$}
\label{sec:nuisance_dm_gamma}

An accurate weak lensing measurement requires an unbiased measurement of the shear using ensemble average of shape information for the source galaxies. Past work has studied the impact of imperfect shape measurements and the resulting residual systematic uncertainties in the $\dSigma$ measurements for the HSC-Y1 sample \citep{2018MNRAS.481.3170M}. To model the impact of possible residual systematic error in the shear calibration, we introduce a nuisance parameter, $\Delta m_\gamma$, and shift the theoretical template as
\begin{align}
\dSigma(R) \rightarrow (1+\Delta m_\gamma)\dSigma(R;\Delta m_\gamma=0). 
\label{eq:dsigma_dm_gamma}
\end{align}
Then we treat $\Delta m_\gamma$ as a nuisance parameter in the cosmological parameter inference and impose a conservative prior as discussed in Section~\ref{sec:bayes}. Since we use a single population of source galaxies as in \citet{Oguri:2010vi}, we can use the same bias parameter $\Delta m_\gamma$ for the lensing signals for the three lens samples (LOWZ, CMASS1, and CMASS2). This is a good approximation as long as the source galaxies are well separated from the lens galaxies. Thus the effect of $\Delta m_\gamma$ does not depend on the lens redshift, so we can distinguish between the two systematic effects of $\Delta z_{\rm ph}$ and $\Delta m_\gamma$, if the signal-to-noise ratio of the $\dSigma$ measurement is sufficiently high. 

When we include both the photo-$z$ errors and the shear multiplicative errors in the parameter inference, we simply multiply the model prediction $\dSigma$ by the multiplicative functions, $1/f_{\rm ph}(\Delta z_{\rm ph})$ and $(1+\Delta m_\gamma)$, assuming that these corrections are independent.

\subsection{Likelihood and parameter estimation}\label{sec:bayes}
We assume that the likelihood of the HSC-Y1 and SDSS galaxy clustering data follows a multivariate Gaussian distribution:
\begin{align}
   -\ln {\cal L}({\bf d}|\boldsymbol{\theta})=
    \frac{1}{2}\left[
    d_i -m_i(\boldsymbol{\theta})
    \right]
    [{\bf C}^{-1}]_{ij}
    \left[
    d_j -m_j(\boldsymbol{\theta})
    \right]
\end{align}
We employ 14 bins in the range $8\le R/[h^{-1}{\rm Mpc}]\le 80$ for $\wgg(R)$ and 8 bins in the range $12\le R/[h^{-1}{\rm Mpc}]\le 80$ for $\dSigma(R)$, respectively. The dimension of the data vector is $66=3\times (14+8)$ for our baseline analysis setup. We use 13 model parameters: 5 cosmological parameters, 3 bias parameters and 3 magnification bias parameters for the three galaxy samples (LOWZ, CMASS1, and CMASS2), and two nuisance parameters ($\Delta z_{\rm ph}$ and $\Delta m_\gamma$). We adopt the 5 cosmological parameters that specify a flat $\Lambda$CDM model in our baseline analysis: $\Omega_{\rm de}$ is the present-day density parameter of the cosmological constant, $\ln (10^{10}A_{\rm s})$ and $n_{\rm s}$ are the amplitude and tilt parameters of the primordial curvature power spectrum normalized at $k_{\rm pivot}=0.05~{\rm Mpc}^{-1}$, $\omega_{\rm b}(\equiv \Omega_{\rm b}h^2)$ is the physical density parameter of baryons,  and $\omega_{\rm c}(\equiv \Omega_{\rm c}h^2)$ is the physical density parameter of CDM. Throughout this paper we employ adiabatic initial conditions.

We use Bayesian parameter inference, where the posterior distribution of the model parameters with a given data vector is expressed as
\begin{align}
    {\cal P}(\boldsymbol{\theta}|{\bf d})\propto {\cal L}({\bf d}|\boldsymbol{\theta})\Pi(\boldsymbol{\theta}),
\end{align}
where $\Pi(\boldsymbol{\theta})$ is the prior on the parameters. The choice of prior is summarized in Table~\ref{tab:parameters}. Since the clustering observables $\dSigma$ and $\wgg$ are not sensitive to $\omega_{\rm b}$ and $n_{\rm s}$, we employ priors on those parameters from other experiments. For the prior on $\omega_{\rm b}$ we employ a normal distribution with mean and width inferred from Big Bang nucleosynthesis (BBN) experiments \citep{Aghanim:2018eyx, Aver:2015iza, Cooke:2017cwo, Schoneberg:2019wmt}. For the prior $n_{\rm s}$ we employ a normal distribution given in Table~\ref{tab:parameters}, where we adopt the {\it Planck} 2018 result \citep[see ``TT,TE,EE+lowE+lensing'' column of Table~2 in Ref.][]{Aghanim:2018eyx} for the mean and a width three times wider than the {\it Planck} error as a conservative choice. For the residual photo-$z$ error parameter, we employ a normal distribution with zero mean and width $\sigma(\Delta z_{\rm ph})=0.1$, which is a conservative choice and larger than the uncertainty inferred from the photo-$z$ method (a few per cent) \citep[see Table.~5 in][]{Hikage:2018qbn}. For the residual shear error parameter, we employ a normal distribution prior with zero mean and width $\sigma(\Delta m_\gamma)=0.01$, which is estimated from HSC galaxy image simulations \cite{2018MNRAS.481.3170M}. For other parameters we adopt a uniform distribution with sufficiently wide width so that these priors are not informative for the cosmological parameter constraints.

We sample the posterior distribution in a multi-dimensional parameter space by using the nested sampling technique implemented in \texttt{MultiNest} \citep{Feroz:2007kg, Feroz:2008xx, Feroz:2013hea} through the interface for cosmological parameter inference \texttt{MontePython} \citep{Audren:2012wb, Brinckmann:2018cvx}. We set the following \texttt{MultiNest} hyper parameters: the sampling efficiency parameter \texttt{efr}=0.8, the evidence tolerance factor \texttt{tol}=0.5, and number of live points \texttt{Nlive}=1000.

\begin{figure}
\begin{center}
    \includegraphics[width=\columnwidth]{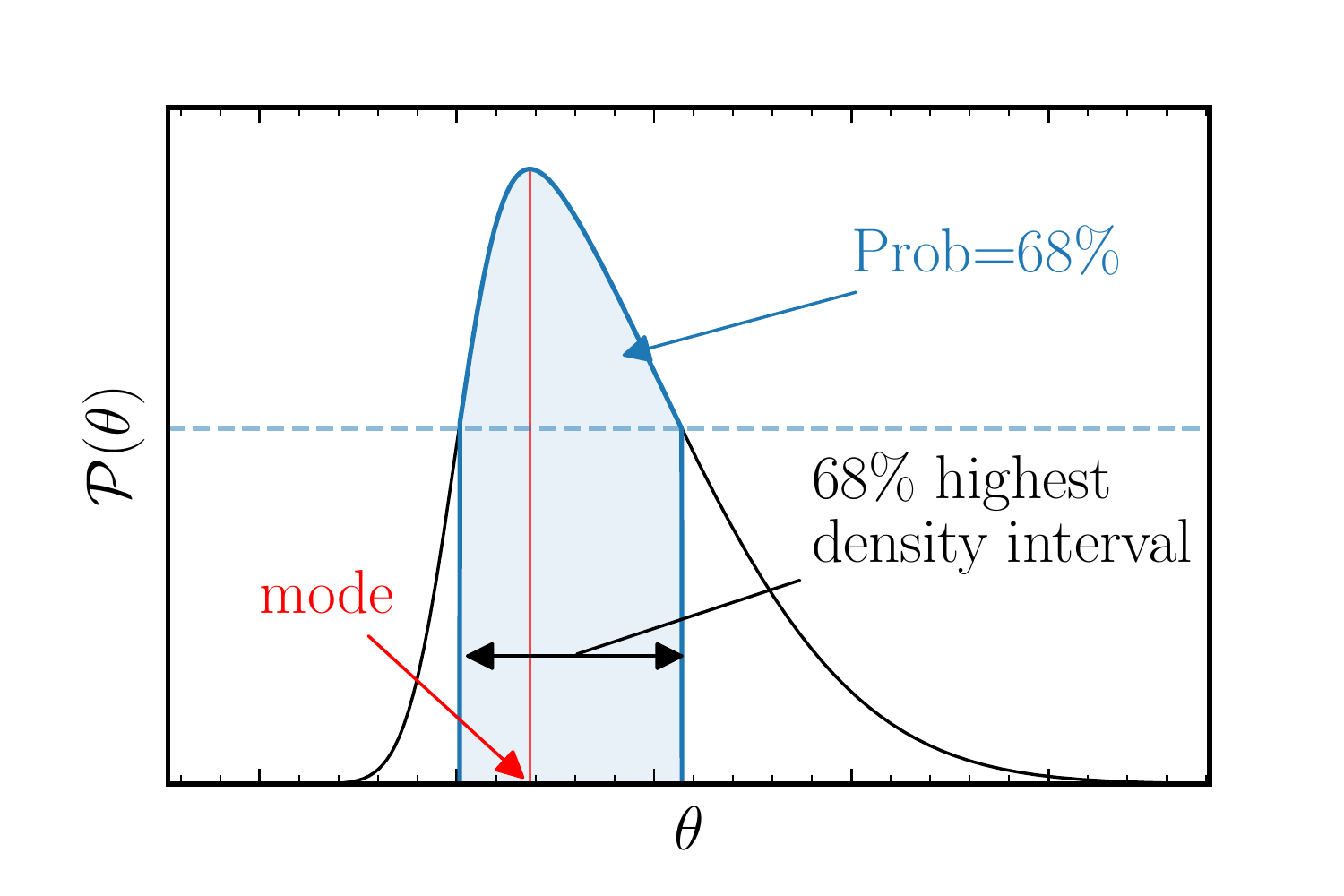}
\end{center}
\caption{An illustration of the definitions of the mode and 68\% highest density interval. The black line is the 1-d marginalized posterior distribution, ${\cal P}(\theta)$, of a model parameter $\theta$. The mode is defined as the parameter value which has highest probability. The 68\% highest density interval is a interval in which the probability is higher than outside and the total probability is 68\%, where probability density is normalized so that $\int_0^\infty {\rm d}\theta{\cal P}(\theta)$=1.}
\label{fig:mhdi-def}
\end{figure}

After sampling the posterior distribution, we estimate the central value and credible interval for the cosmological parameter(s) from the chain. In this paper, we report the mode as the central value and the 68\% highest density interval  as the credible interval. The definitions of the mode and the highest density interval are illustrated in Fig.~\ref{fig:mhdi-def}. The mode is defined by the parameter value that has the highest posterior probability, and the 68\% highest density interval is defined so that the probability integrated within the interval is 68\%. To obtain the mode and the highest density interval from a given chain, we first estimate the marginalized posterior distribution by using kernel density estimation using the public code {\tt getdist} \citep{Lewis:2019xzd}, and then find the mode and the highest density interval from the  posterior distribution.

Among the sampled model parameters, we focus on $\Omega_{\rm m}$, $\sigma_8$, and $S_8\equiv\sigma_8(\Omega_{\rm m}/0.3)^{0.5}$, which are derived from the sampled parameters. In Appendix~\ref{sec:nestcheck}, we show the convergence of our sample chains and that our setup is suitable to infer the cosmological parameters to within a sub-percent level of the 68\% credible interval. 

\section{Blinding scheme and Validation}
\label{sec:blinding}
In this section, we describe our strategy of analysis to secure the robustness of the results to confirmation bias and systematics in the data, analysis and model. 

To avoid confirmation bias we perform our cosmological analysis in a blind fashion. The details are described in Section~3.2 of \citet{Hikage:2018qbn}. We employ a two-tier blinding strategy to avoid accidental unblinding during the cosmological analysis; namely, both at the catalog level and at the analysis level:
\begin{itemize}
\item {\it catalog level} -- the analysis team performs the full cosmological analysis using three different catalogs of galaxy shapes for weak lensing measurements, one of which is the true catalog and the other two are fake (see below for details). No member in the analysis team knows which catalog is the true catalog. 
\item {\it analysis level} -- the analysis team is not allowed to make a plot comparing the measurement with theoretical models. When the analysis team makes a plot showing the credible regions of cosmological parameters (i.e. the posterior distribution), the central value(s) of parameter(s) are shifted to zero, and only the shifted range of the credible region(s) can be seen. 
\item {\it analysis level} -- The analysis team is not allowed to compare the posterior of cosmological parameter(s) or the inferred model predictions with external results such as the {\it Planck} CMB cosmology. 
\end{itemize}

Use of the three shape catalogs means that the analysis group must perform three analyses, but this method avoids the need for reanalyses once the catalogs are unblinded.

In addition we did not make any comparison between the posterior distributions of parameters obtained from this paper and from the companion paper using the halo model method, \citet{Miyatake:2021b}, during the blind analysis stage.  Our cosmological analysis method has been tested and demonstrated to be valid in \citet{Sugiyama:2020kfr}.

In addition to our baseline analysis, we employ alternative analysis setups to quantify internal consistency and systematic effects in parameter estimation {\it before} unblinding, as summarized in Table~\ref{tab:analsyis-setup}. 

{\renewcommand{\arraystretch}{1.2} 
\begin{table*}
\caption{The list of our analysis setups: the baseline analysis setup and those designed to check for internal consistency. The marks ``\yes'' or ``\no'' in each column denote whether each analysis does or does not include the parameter(s) in the inference. The column labelled as ${\boldsymbol{\theta}}_{\rm cosmo}$ denotes the set of cosmological parameters, $\boldsymbol{\theta}_{\rm cosmo}=\{\Omega_{\rm de},\ln(10^{10}A_{\rm s}),n_{\rm s},\omega_{\rm c},\omega_{\rm b}\}$. The other parameters are the same as in Table~\ref{tab:parameters}.
}
\begin{center}
\begin{tabular}{l|cccccc|l}
\hline\hline
setup label & \multicolumn{6}{c}{sample parameter} & comment \\
& $\boldsymbol{\theta}_{\rm cosmo}$ & $(b_{1,i})$ & $\Delta m_{\gamma}$ & $\Delta z_{\rm ph}$ & $\alpha_{\rm mag}$ & $w$ & \\
\hline
baseline & \yes& \yes& \yes&\yes& \yes& \no& baseline analysis: $\boldsymbol{\theta}_{\rm cosmo}=\{\Omega_{\rm de},\ln(10^{10}A_{\rm s}),n_{\rm s},\omega_{\rm c},\omega_{\rm b}\}$\\ 
w/o LOWZ   & \yes& \yes& \yes& \yes&\yes& \no& w/o the LOWZ sample\\
w/o CMASS1 & \yes& \yes& \yes& \yes&\yes& \no& w/o the CMASS1 sample \\
w/o CMASS2 & \yes& \yes& \yes& \yes&\yes& \no& w/o the CMASS2 sample\\
w/o photo-$z$ error   & \yes& \yes& \yes& \no& \yes & \no& fixing $\Delta z_{\rm ph}=0$\\
w/o shear calib. error& \yes& \yes& \no & \yes& \yes& \no& fixing $\Delta m_{\gamma}=0$\\
w/o mag. effect error & \yes& \yes& \yes& \yes& \no & \no& fixing $\alpha_{\rm mag}=0$\\
2 cosmo. paras & $\triangle$ &\yes& \yes& \yes & \yes& \no& fixing $(n_{\rm s},\omega_{\rm b}, \omega_{\rm c})$ to the {\it Planck} best-fit values: $\boldsymbol{\theta}_{\rm cosmo}=\{\Omega_{\rm de},\ln(10^{10}A_{\rm s})\}$\\
\hline
$w$CDM & \yes& \yes& \yes& \yes& \yes& \yes& including the DE equation of state parameter, 
$w_{\rm de}$, {\it after} unblinding \\
\hline\hline
\end{tabular}
\label{tab:analsyis-setup}
\end{center}
\end{table*}
}

\section{Results}
\label{sec:results}

\subsection{Cosmological parameters for the $\Lambda$CDM model}
\label{sec:cosmology-constraint}

In this section, we present the results of cosmological constraints in a flat $\Lambda$CDM model. All the analysis we show in this section were done before unblinding \footnote{After unblinding, we found a bug in the analysis pipeline that the width of the Gaussian prior was mistakenly set to a value smaller than the correct value by a factor of $2^{1/2}$. The results in this section were from the reanalyses with this bug fixed, done after unblinding. However, we believe that the blinding strategy was still effective and our results are free from confirmation bias, because the results before and after the bug fixed were almost unchanged; therefore we concluded that the unblinding criteria are not affected.}. We neither saw the actual values of the cosmological parameters, nor compared our results with those from any other analysis, including those from {\it Planck} or our companion paper, \citet{Miyatake:2021b}.

{\renewcommand{\arraystretch}{1.3} 
\begin{table*}
\caption{Summary of cosmological constraints for different analysis setups (see Table~\ref{tab:analsyis-setup}). Here we report the mode of the marginalized posterior along with the credible interval defined by the highest density interval (see Section~\ref{sec:bayes} and Fig.~\ref{fig:mhdi-def}).}
\label{tab:summary-table}
\begin{center}
\begin{tabular}{l|cccc}
\hline\hline
setup & $S_8\equiv\sigma_8(\Omega_\mathrm{m}/0.3)^{0.5}$ & $\sigma_{8 } $ & $\Omega_{\rm m}$ & $w_0$\\
\hline
baseline & $0.936^{+0.092}_{-0.086}$ & $0.85^{+0.16}_{-0.11}$ & $0.283^{+0.12}_{-0.035}$ & $-$\\
w/o LOWZ & $1.13^{+0.14}_{-0.13}$ & $1.02^{+0.17}_{-0.15}$ & $0.328^{+0.12}_{-0.054}$ & $-$\\
w/o CMASS1 & $0.99^{+0.10}_{-0.12}$ & $0.94^{+0.19}_{-0.15}$ & $0.255^{+0.11}_{-0.037}$ & $-$\\
w/o CMASS2 & $0.92^{+0.11}_{-0.12}$ & $0.82^{+0.16}_{-0.14}$ & $0.319^{+0.10}_{-0.052}$ & $-$\\
w/o multiplicative bias & $0.932^{+0.095}_{-0.089}$ & $0.84^{+0.17}_{-0.11}$ & $0.288^{+0.12}_{-0.037}$ & $-$\\
w/o photo-$z$ & $0.941^{+0.078}_{-0.082}$ & $0.85^{+0.15}_{-0.10}$ & $0.298^{+0.11}_{-0.042}$ & $-$\\
w/o mag. effect & $0.939^{+0.088}_{-0.087}$ & $0.86^{+0.14}_{-0.12}$ & $0.287^{+0.12}_{-0.034}$ & $-$\\
DEmP & $0.931^{+0.083}_{-0.089}$ & $0.85^{+0.18}_{-0.11}$ & $0.270^{+0.11}_{-0.035}$ & $-$\\
Ephor AB & $1.071^{+0.077}_{-0.082}$ & $1.07^{+0.14}_{-0.16}$ & $0.260^{+0.090}_{-0.041}$ & $-$\\
Franken-Z & $0.948^{+0.086}_{-0.087}$ & $0.86^{+0.17}_{-0.11}$ & $0.278^{+0.11}_{-0.037}$ & $-$\\
Mizuki & $0.892^{+0.089}_{-0.085}$ & $0.79^{+0.17}_{-0.10}$ & $0.28^{+0.13}_{-0.030}$ & $-$\\
NNPZ & $0.952^{+0.092}_{-0.088}$ & $0.86^{+0.17}_{-0.11}$ & $0.288^{+0.12}_{-0.036}$ & $-$\\
2 cosmo & $0.939^{+0.090}_{-0.087}$ & $0.909^{+0.10}_{-0.089}$ & $0.311^{+0.036}_{-0.037}$ & $-$\\
$w$CDM (HSC$\times$Planck2018) & $0.817^{+0.022}_{-0.021}$ & $0.892^{+0.051}_{-0.056}$ & $0.246^{+0.045}_{-0.035}$ & $-1.28^{+0.20}_{-0.19}$\\
\hline\hline
\end{tabular}
\end{center}
\end{table*}
}
%
\begin{figure}
\begin{center}
    \includegraphics[width=\columnwidth]{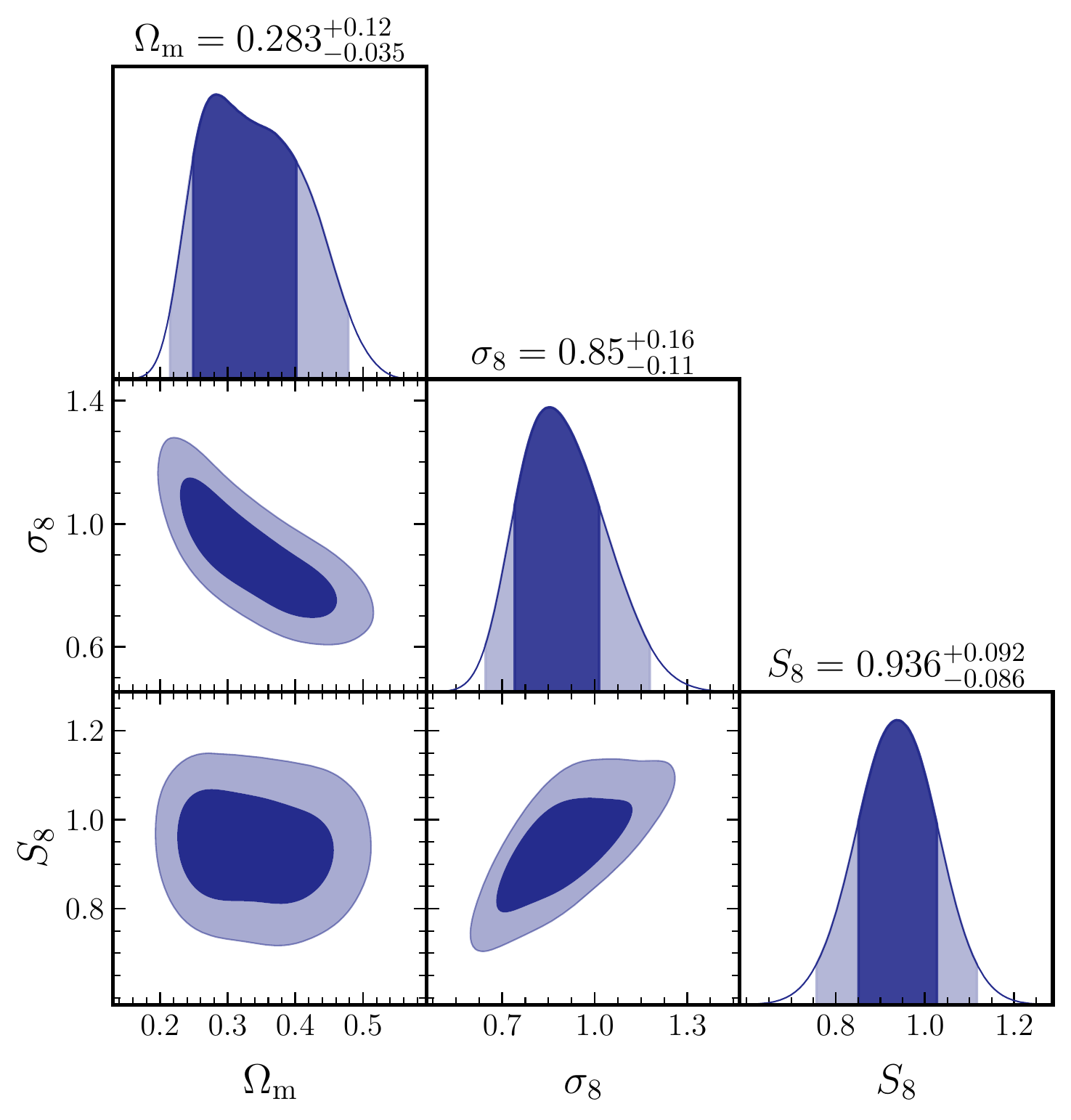}
\end{center}
\caption{The 1-d and 2-d posterior distributions of the three main cosmological parameters, $\Omega_{\rm m}$, $\sigma_8$ and $S_8$, obtained from the baseline analysis (see Table~\ref{tab:analsyis-setup}), where we use the galaxy clustering signals over $8<R/[h^{-1}{\rm Mpc}]<80$ and the galaxy-galaxy lensing signals over $12<R/[h^{-1}{\rm Mpc}]<80$. The dark (light) shaded region shows the 68\% (95\%) credible interval. The posterior distributions over the full 13-parameter space are shown in Fig.~\ref{fig:corner-baseline-full}.}
\label{fig:corner-baseline}
\end{figure}

\begin{figure}
\begin{center}
    \includegraphics[width=\columnwidth]{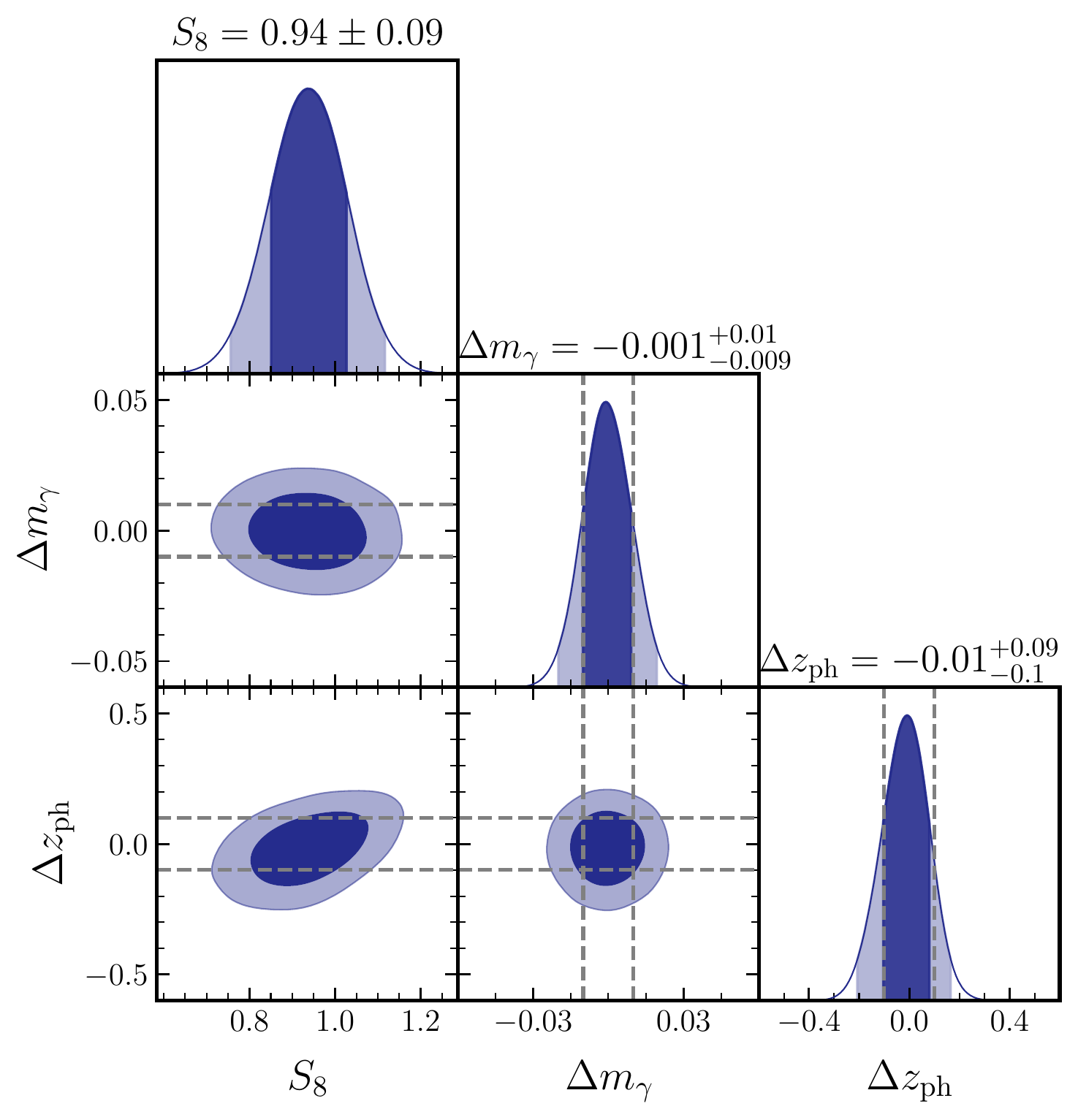}
\end{center}
\caption{Similar to the previous figure, now showing the 1-d and 2-d posterior distributions for $S_8$ and the nuisance parameters $\Delta m_\gamma$ (multiplicative shear error) and $\Delta z_{\rm ph}$ (photo-$z$ bias) for the HSC source galaxies used in the weak lensing measurements. The vertical dashed lines in the 1-d distributions of $\Delta m_\gamma$ and $\Delta z_{\rm ph}$ denote the width of the Gaussian prior of these parameters.}
\label{fig:corner-baseline-nuisance-self-calibration}
\end{figure}

Fig.~\ref{fig:corner-baseline} shows the posterior distributions of $S_8$, $\sigma_8$ and $\Omega_{\rm m}$ for the baseline analysis. The central value and the credible interval of each parameter are, respectively:
\begin{equation}
\begin{aligned}
     S_8&=0.936^{+0.092}_{-0.086}\\
     \sigma_8&=0.85^{+0.16}_{-0.11}\\
     \Omega_{\rm m}&=0.283^{+0.12}_{-0.035}
\end{aligned}
\end{equation}
Thus the joint measurements of $\dSigma$ and $\wgg$ from the HSC-Y1 and SDSS catalogs achieve a precision of $\sigma(S_8)\simeq 0.09$. This precision can be considered as a conservative and robust constraint, because we demonstrated in \citet{Sugiyama:2020kfr} that any systematic bias in $S_8$ due to inaccuracies in the minimal bias model is very unlikely to become larger than the 68\% credible interval. 
However, our result displays a slightly larger value of $S_8$ compared to other results such as the {\it Planck} result, as we below discuss in more detail. For completeness, we show the posterior distributions of all 13 parameters (and several derived parameters) in Appendix~\ref{sec:full-corner-plot}. A closer look at Fig.~\ref{fig:corner-baseline} shows that the posterior distribution of $\Omega_{\rm m}$ has a flat-shaped peak. We found that this arises when we combine the cosmological information from the three galaxy samples, while  the cosmological parameters from each of the galaxy samples alone differ due to sample variance, as shown in Appendix~\ref{sec:Om-bimodal} \citep[also see][]{2020JCAP...05..042I}. The flat-shaped peak is also partially a consequence of the degeneracy between $\Omega_{\rm m}$ and $\omega_{\rm c}$. If we fix $\omega_{\rm c}$ to the {\it Planck} 2015 best-fit value, $\Omega_m$ shows a narrower and peakier distribution.

As discussed in detail in \citet{Sugiyama:2020kfr}, using either of $\dSigma(R)$ or $\wgg(R)$ alone cannot constrain these parameters simultaneously and suffers from severe degeneracies. On large scales where linear theory holds, $\dSigma(R)$ is proportional to $b_1 \sigma_8^2$, while $\wgg(R)$ is proportional to $b_1^2 \sigma_8^2$. Thus either alone cannot constrain $b_1$ and $\sigma_8$ separately. If $\sigma_8$ is very large, that assumption is incorrect, and our model will overpredict $\dSigma$ and $\wgg$ around the minimum scale because the stronger nonlinear effect boosts the amplitudes around the scale compared to what the linear theory predicts. Because of this, a model with extremely large $\sigma_8$ is disfavored in parameter inference, while there is no such penalty for arbitrarily small values of $\sigma_8$. Thus the resultant posterior distribution with either $\dSigma$ or $\wgg$ alone depends on the lower limit of the prior range of $\sigma_8$. Hence only the joint analysis of $\dSigma$ and $\wgg$ gives meaningful constraints on $\sigma_8$ and  $S_8$ \citep[see also Fig.~7 of Ref.][]{Sugiyama:2020kfr}. As shown in Appendix~\ref{sec:full-corner-plot}, the bias parameter for each galaxy sample is determined to a fractional precision of about 20\%. $\Omega_{\rm m}$ is constrained relatively well because a change in $\Omega_{\rm m}$ causes a scale-dependent modification in $\wgg$ (and $\dSigma$), so the $\wgg$ information in our datasets can give a meaningful constraint on the parameters. 

Fig.~\ref{fig:corner-baseline-nuisance-self-calibration} shows the posterior distributions of $S_8$ with the nuisance parameters characterizing photo-$z$ errors and multiplicative shear bias. The posterior distributions of the nuisance parameters are prior-dominated (see Table~\ref{tab:parameters}), meaning that  $\dSigma(R)$ and $\wgg(R)$ on large scales alone cannot constrain these parameters well. As shown in \citet{Miyatake:2021b} the inclusion of smaller-scale information in the analysis constrains the photo-$z$ parameter better than the prior range. In addition,
\citet{Miyatake:2021b} presented the cosmological parameters when employing an even broader width of $\sigma(\Delta z_{\rm ph})=0.2$, instead of $\sigma(\Delta z_{\rm ph})=0.1$ in the fiducial prior, and showed that the data allows for a self-calibration of the residual photo-$z$ bias, indicating $\Delta z_{\rm ph}=-0.113$ for the central value, as can be found from Fig.~26 of their paper. As the post-unblinding analysis, we found that $S_8$ is changed to $S_8=0.902^{+0.077}_{-0.082}$ from $S_8=0.936^{+0.092}_{-0.086}$, if we adopt the fixed value of $\Delta z_{\rm ph}=-0.113$ in the parameter inference. Thus the larger value of $S_8$ in our results might be partly from the possible residual photo-$z$ bias. However, the shift is not significant, and we need more HSC data to obtain a more definite conclusion.

\begin{figure}
\begin{center}
    \includegraphics[width=\columnwidth]{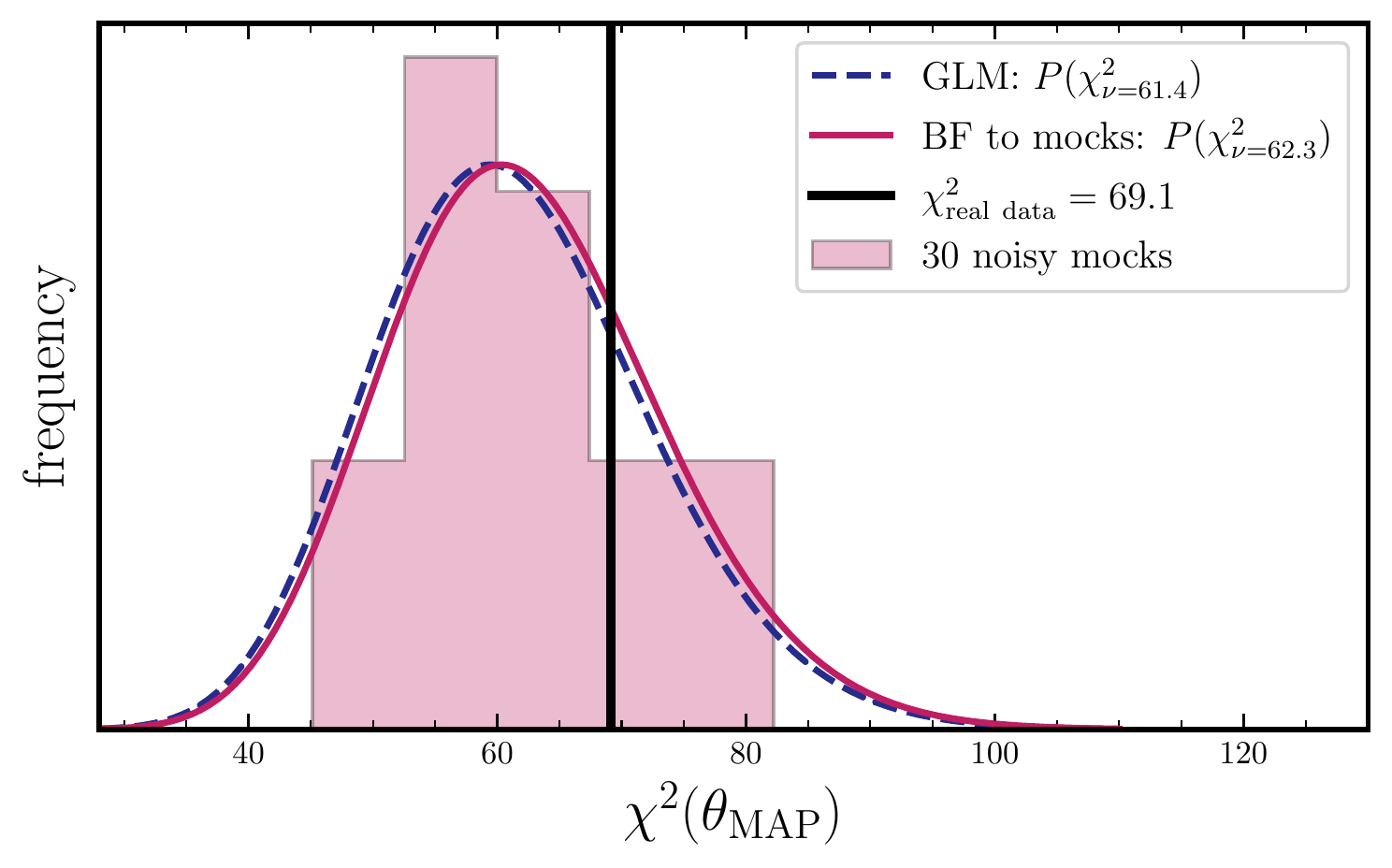}
\end{center}
\caption{An evaluation of goodness-of-fit of the best-fit model at {\it maximum a posteriori} (MAP) for the baseline analysis. The magenta histogram shows the distribution of the $\chi^2$ value of the model at MAP, obtained by applying the same baseline analysis to each of 30 noisy mock datasets (see text for details). The magenta line is the best-fit $\chi^2$ distribution, characterized by the degrees of freedom $\nu=62.3$ estimated from the same mocks. The vertical black line indicates $\chi^2=69.1$ at MAP for the real analysis of the HSC-Y1 and SDSS data. The blue dashed line shows the $\chi^2$ distribution with degrees of freedom $\nu=60.5$ computed using the Gaussian linear model (GLM) in \citet{2020PhRvD.101j3527R} (see text for details).}
\label{fig:chi-square}
\end{figure}
Fig.~\ref{fig:chi-square} shows an evaluation of goodness-of-fit of the best-fit model, evaluated at {\it maximum a posteriori} (MAP) of the posterior distribution, for the baseline analysis. To make the reference $\chi^2$ distribution, we need to evaluate the effective degrees of freedom, which generally differ from the naive evaluation of degrees of freedom, i.e. $\nu=\nu_{\rm data}-\nu_{\rm param}=66-13=53$ in our case, because of the parameter degeneracies. To obtain the effective degrees of freedom, we first generate noisy mock data vectors of $\dSigma(R)$ and $\wgg(R)$ for the galaxy samples by adding random statistical scatter to the {\it noiseless} mock signals of $\dSigma(R)$ and $\wgg(R)$. Note that the noiseless mock signals were generated using $N$-body simulation data for the {\it Planck} 2015 cosmology \cite{Sugiyama:2020kfr} \citep[also see][]{Nishimichi:2018etk}. The histogram in the figure shows the distribution of $\chi^2$ for the model at MAP for each of the 30 noisy mocks. The magenta line denotes the best-fit $\chi^2$-distribution, specified by the effective degrees of freedom $\nu_{\rm eff}=62.3$. The actual $\chi^2$ at MAP for the real analysis of HSC-Y1 and SDSS data is $\chi^2=69.1$, corresponding to a $p$-value of $0.259$, which indicates that the minimal bias model gives an acceptable fit to the data within the error bars. For a further comparison, we also compute the expected $\chi^2$ distribution following the ``Gaussian linear model'' (GLM) in Ref.~\cite{2020PhRvD.101j3527R}, which assumes that both data vector and model parameters are Gaussian-distributed. Here we generate 10,000 noisy mock data vectors around the above mock signals, compute the MAP model for each of them, and compute $\chi^2$ at MAP using the GLM method. The blue dashed line gives the best-fit $\chi^2$-distribution, specified by the effective degrees of freedom $\nu_{\rm eff}=60.5$ estimated from these 10,000 mocks. The GLM method gives a consistent $\chi^2$ distribution with that estimated from the analysis using the noisy mock signals. Hence we conclude the MAP model gives an acceptable fit to the measured signals.

\begin{figure*}
\begin{center}
    \includegraphics[width=1.8\columnwidth]{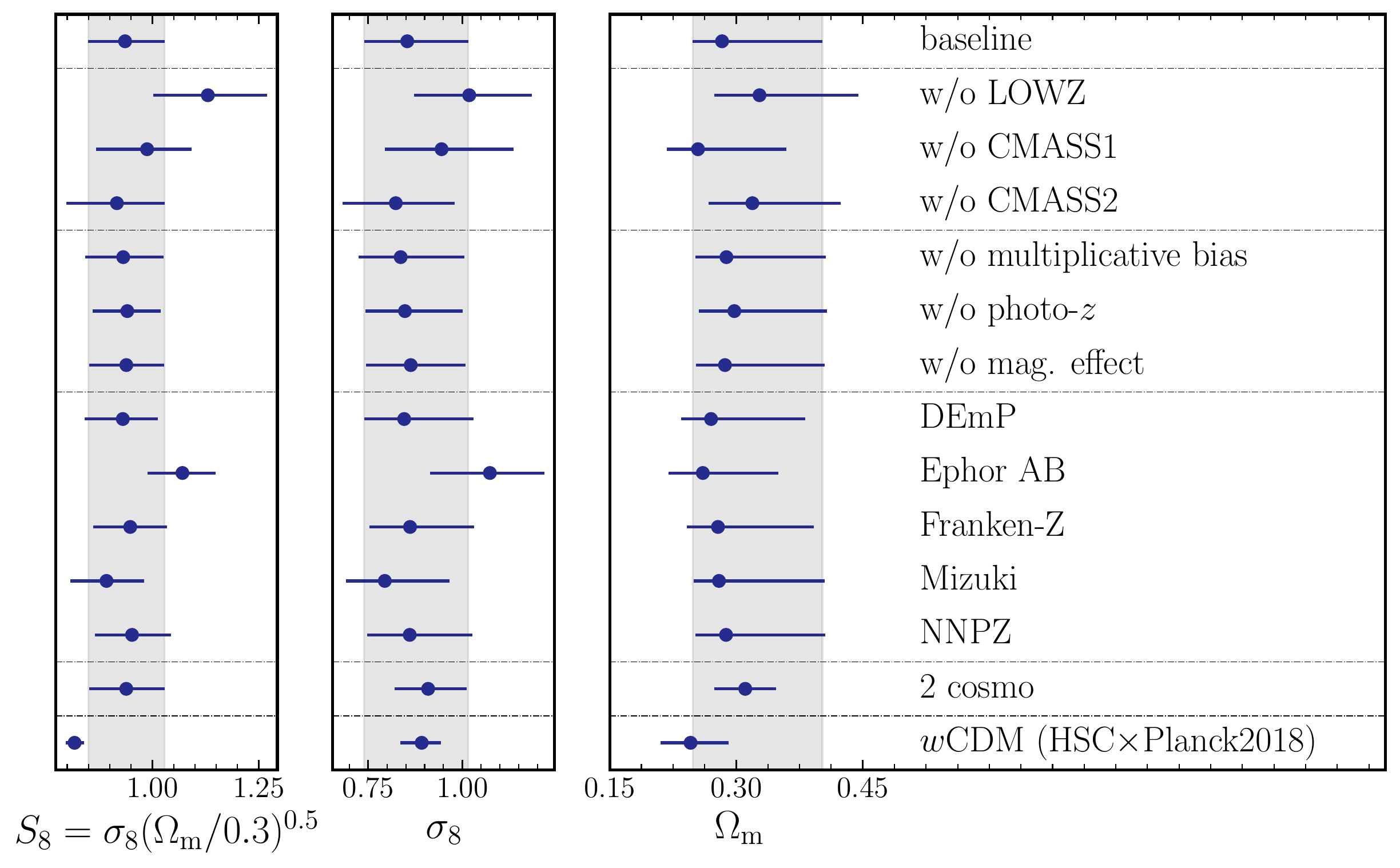}
\end{center}
\caption{Summary of the constraints on cosmological parameters for the different setups (see Table~\ref{tab:analsyis-setup}). Here the dot symbol and error bar in each column denote the central value and the 68\% credible interval (Fig.~\ref{fig:mhdi-def}). For comparison, the shaded band shows the 68\% credible interval of the baseline analysis.}
\label{fig:bar-plot}
\end{figure*}
%
Fig.~\ref{fig:bar-plot} summarizes the 1-d posterior distribution of $S_8$, $\sigma_8$ and $\Omega_{\rm m}$ for the different setups in Table~\ref{tab:analsyis-setup}. We do not identify any significant shift or sign of systematic effects in the parameter estimation; in particular, all the $S_8$ results are consistent with the baseline setup to within the 68\% credible interval. In the following we discuss each of the results.

\begin{figure}
\begin{center}
    \includegraphics[width=\columnwidth]{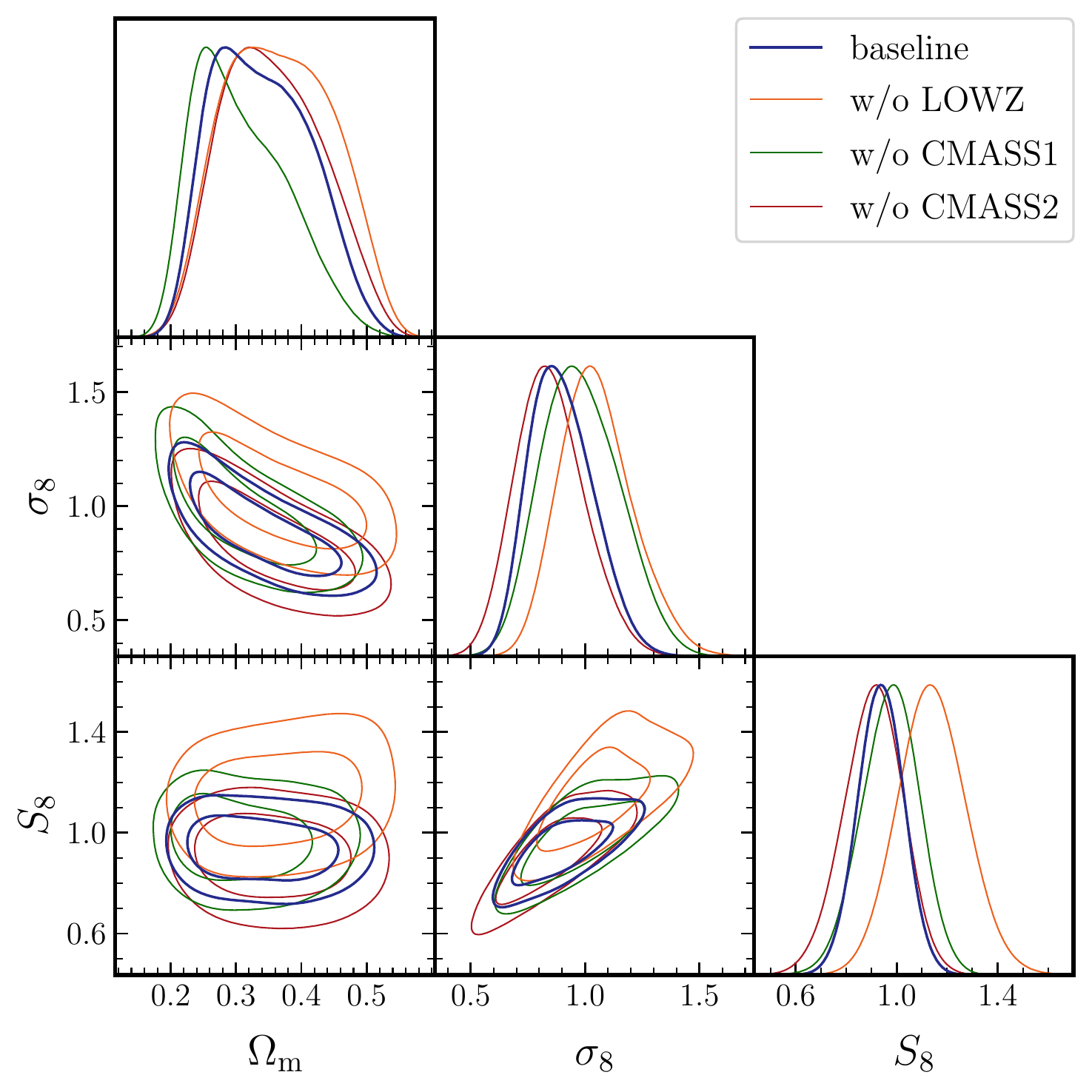}
\end{center}
\caption{Similar to Fig.~\ref{fig:corner-baseline}, now showing the posterior distributions when we remove one of the galaxy samples (LOWZ, CMASS1 or CMASS2) from the parameter inference. }
\label{fig:corner-wo-galsample}
\end{figure}

\begin{figure}
\begin{center}
    \includegraphics[width=\columnwidth]{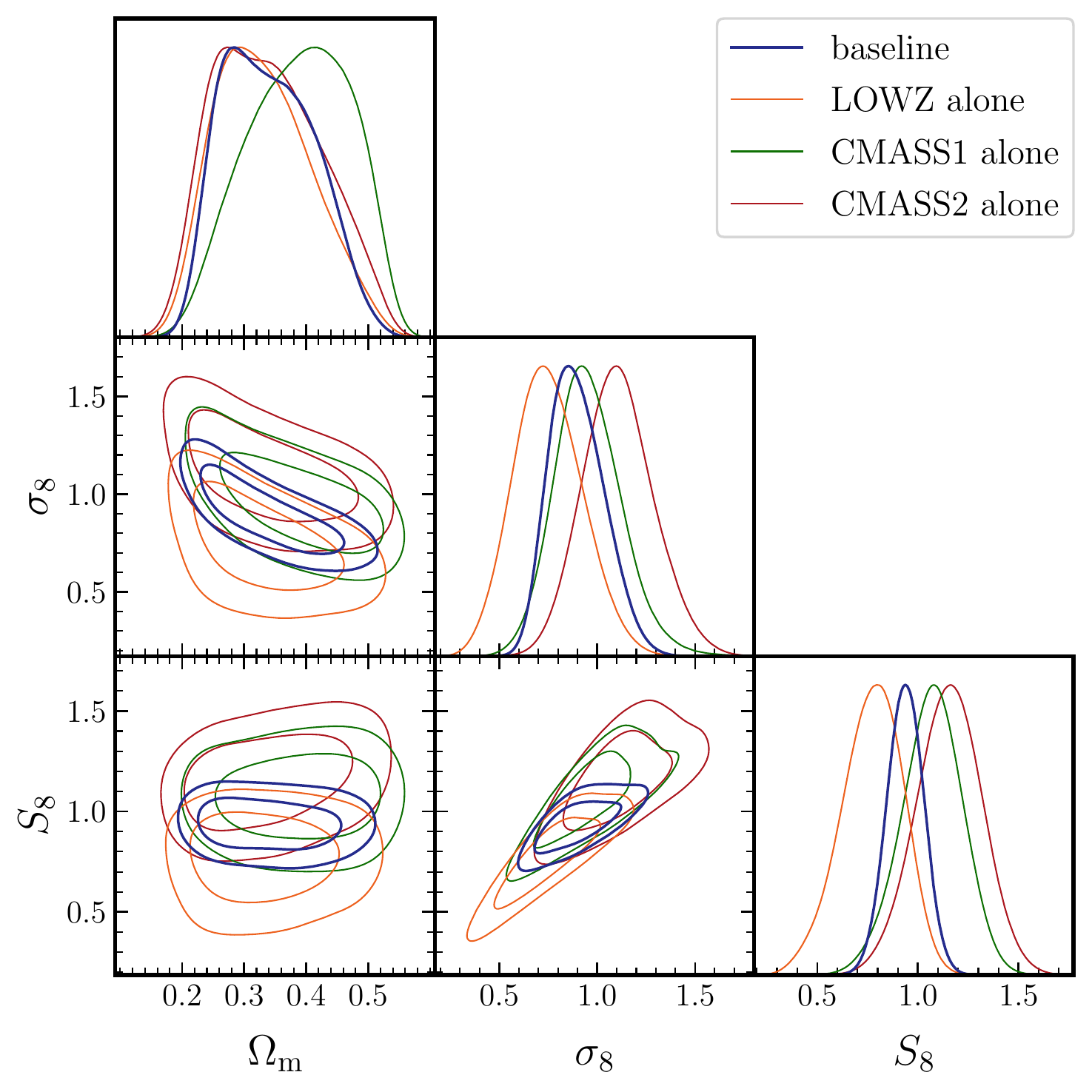}
\end{center}
\caption{Similar to Fig.~\ref{fig:corner-baseline}, now showing the posterior distributions for each galaxy sample alone.}
\label{fig:corner-galsample_alones}
\end{figure}

%
\begin{figure}
\begin{center}
    \includegraphics[width=\columnwidth]{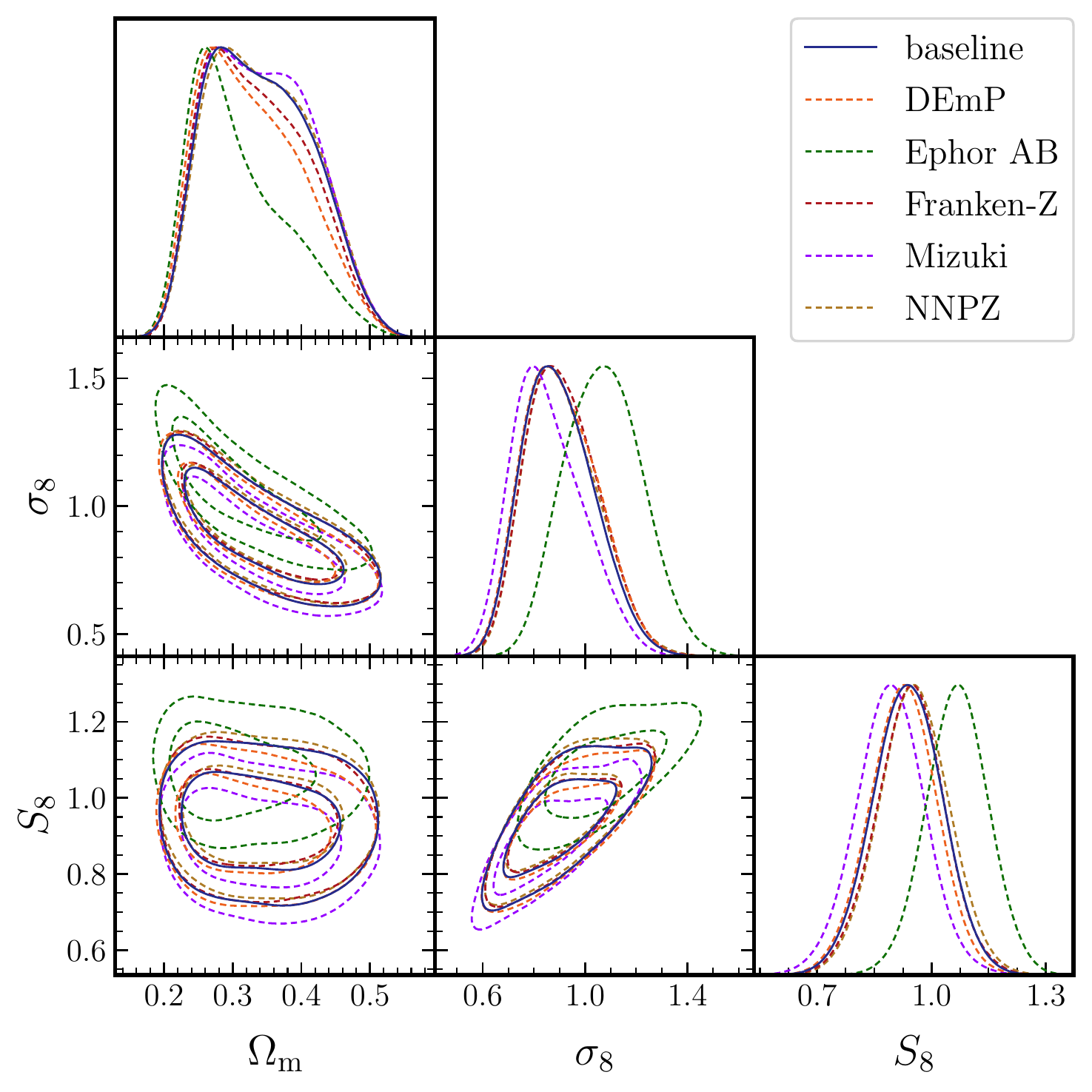}
\end{center}
\caption{Similar to Fig.~\ref{fig:corner-baseline}, now showing  the results when we use different photo-$z$ catalogs used to define the HSC source galaxies for the $\dSigma$ measurements (see text for details).}
\label{fig:corner-photoz}
\end{figure}
%
Fig.~\ref{fig:corner-wo-galsample} shows the results if we remove one of the three galaxy samples (LOWZ, CMASS1 or CMASS2) from the parameter inference. In Fig.~\ref{fig:corner-galsample_alones} we show the posterior distributions for each galaxy sample separately.  In both cases, the results are consistent with one another. These plots demonstrate that the scatter in the posterior distribution has an appreciable contribution from sample variance.  

In Fig.~\ref{fig:corner-photoz} we study how the inferred cosmological parameters are changed if we use the different photo-$z$ catalogs of HSC source galaxies based on Eq.~(\ref{eq:def_source_galaxies}) for the $\dSigma$ measurements. We used the same covariance matrix as for the fiducial photo-$z$ catalog for these analyses, because the effect of different photo-$z$ methods on the covariance matrix was shown to be small by Ref.~\cite{2019MNRAS.486...52S}. Keeping the covariance matrix fixed also allows a more direct comparison of the sensitivity of cosmological parameters to the choice of photo-$z$ catalog. We find consistent results within the credible intervals for all photo-$z$ samples except for ``Ephor~AB'', which differs significantly from the results of the baseline analysis. The Ephor-AB sample has a substantially smaller number of source-lens pairs, giving a noisier $\dSigma$ signal. We explore this further in Appendix~\ref{sec:ephor-test}, where we find that the difference in $\sigma_8$ and $S_8$ is caused largely by the CMASS2 signal at large separations, $R>10h^{-1}{\rm Mpc}$; if we remove the CMASS2 sample, we find much more consistent results. 

\begin{figure}
\begin{center}
    \includegraphics[width=\columnwidth]{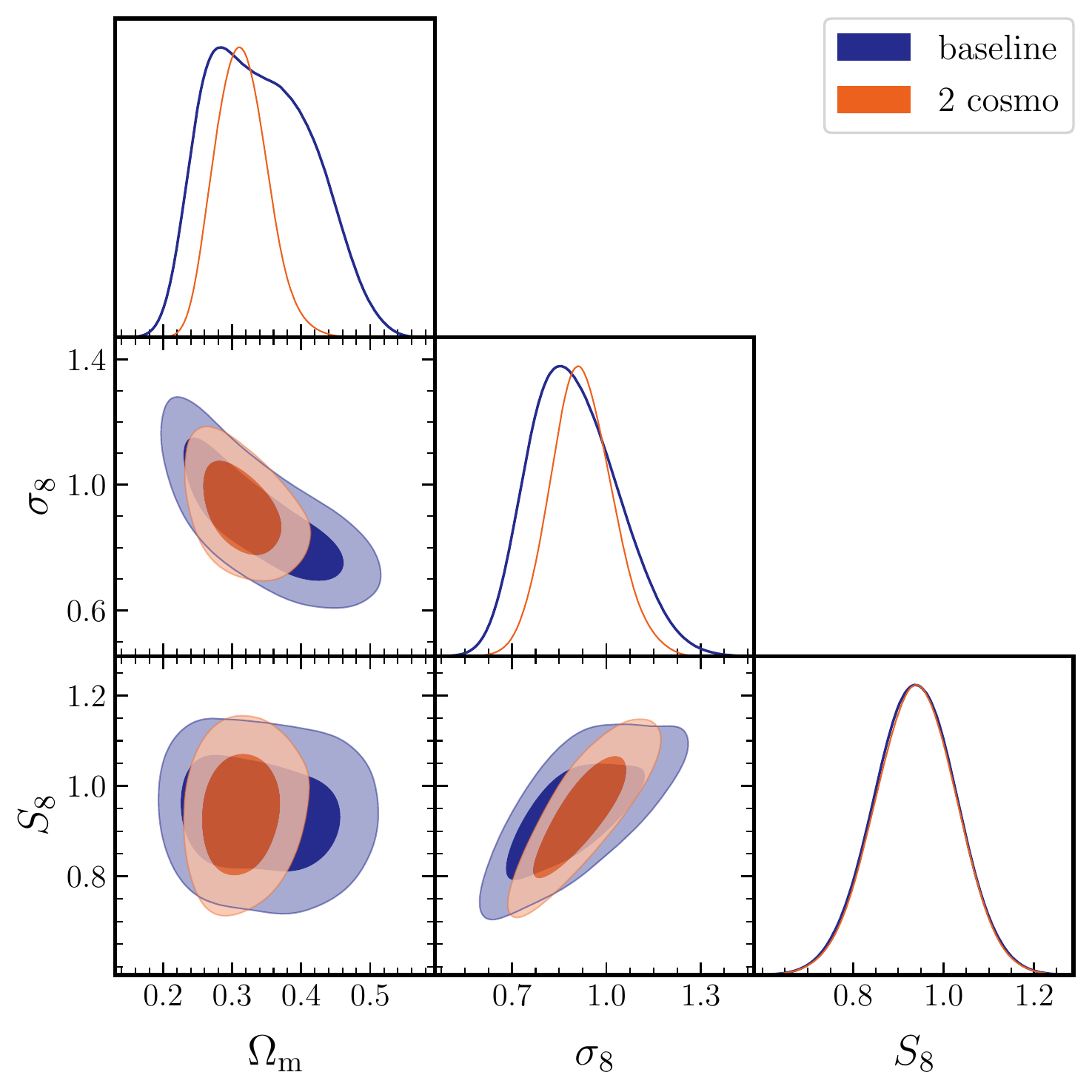}
\end{center}
\caption{The posterior distributions when using only the two cosmological parameters, $\Omega_{\rm de}$ and $\ln(10^{10}A_{\rm s})$, in the parameter inference, instead of the five parameters in the baseline analysis in Fig.~\ref{fig:corner-baseline}. We fixed the values of the other three cosmological parameters to those inferred from the {\it Planck} 2015 ``TT,TE,EE+lowP'' constraints, and we included the same nuisance parameters as in the baseline analysis. The $S_8$ result is almost unchanged.}
\label{fig:corner-2vs5cosmo}
\end{figure}
%
Fig.~\ref{fig:corner-2vs5cosmo} shows the results when allowing only the two cosmological parameters, $\Omega_{\rm de}$ and $\ln(10^{10}A_{\rm s})$ to vary, while keeping the other three parameters, $(\omega_{\rm b}, \omega_{\rm c}, n_{\rm s})$ to the best-fit values of the {\it Planck} 2015 ``TT,TE,EE+lowP'' constraints \citep{Aghanim:2018eyx}. Encouragingly the $S_8$ result is almost unchanged, confirming that $S_8$ is close to the principal parameter that the clustering observables can most accurately constrain. In n other words, other three parameters do not strongly affect the inference from clustering observables. On the other hand, the $\Omega_{\rm m}$ constraint is considerably weaker when allowing all five parameters to vary, because of the strong degeneracy between $\Omega_{\rm m}$ and $\omega_{\rm c}$ in a flat $\Lambda$CDM model. Since $\sigma_8$ and $\Omega_{\rm m}$ are correlated in our analysis (Fig.~\ref{fig:corner-baseline-full}), this leads to a degradation in the constraints on $\sigma_8$.

\subsection{Cosmological results of post-unblinded analyses}
\label{sec:post-unblind}

In this section, we show the results of post-unblinding analyses. 

\begin{figure}
    \centering
    \includegraphics[width=\columnwidth]{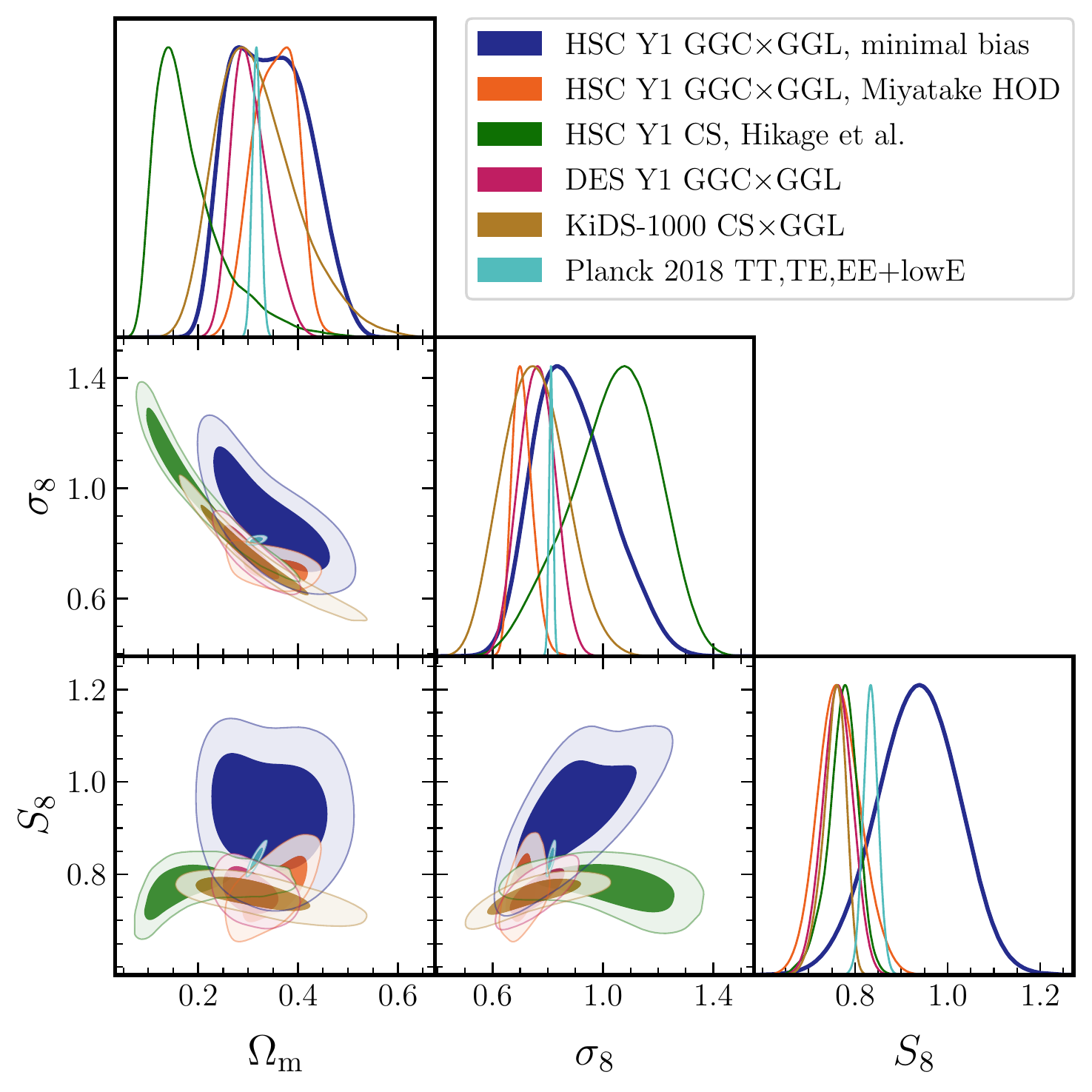}
    \caption{Comparison of our constraints (``HSC-Y1 GCC$\times$GGL, minimal bias'') with other cosmological experiments for flat-geometry $\Lambda$CDM cosmologies. For the galaxy survey constraints, we focus on the results that are obtained from clustering information using a similar setup to ours: ``GGC'' denotes the projected galaxy clustering, ``GGL'' denotes the galaxy-galaxy weak lensing, and ``CS'' denotes the cosmic shear information. Magenta contours show the DES-Y1 results \cite{2018PhRvD..98d3526A}, while brown contours show the KiDS-1000 result \citep{2021A&A...646A.140H}. The orange contours show the results obtained in our companion paper, \citet{Miyatake:2021b}, when including smaller scale information with the HOD based method for the same HSC-Y1 data, ``HSC-Y1 GCC$\times$GGL, Miyatake HOD''. The green contours show the results for cosmic shear tomography, ``HSC-Y1 GC, Hikage'' \citep{Hikage:2018qbn}. The cyan contours show the {\it Planck} CMB constraint.}
    \label{fig:corner-comparion-externals}
\end{figure}

In Fig.~\ref{fig:corner-comparion-externals}, we compare cosmological parameter estimation results with those from other experiments for the flat $\Lambda$CDM model. Throughout, we assumed a fixed value of total neutrino mass, $m_{\nu,{\rm tot}}=0.06\,$eV. For the {\it Planck} result, we consider the {\it Planck} 2018 \citep{Aghanim:2018eyx} cosmological constraints, in particular those derived from primary CMB information, referred to as ``TT, EE, TE+lowE'' in their paper. In other words, we do not include CMB lensing information. We used the publicly available likelihood code for the {\it Planck} data \footnote{\url{https://wiki.cosmos.esa.int/planck-legacy-archive/index.php/Cosmological_Parameters}} to infer the cosmological parameters when the neutrino mass is fiexd to 0.06~eV. The HSC-Y1 cosmic shear result \citep{Hikage:2018qbn} is taken from the website\footnote{\url{https://hsc-release.mtk.nao.ac.jp/archive/filetree/s16a-shape-catalog/pdr1_hscwl/Hikage/HSC_Y1_LCDM_post_mnu0.06eV.txt}}. For the other lensing experiments that we compare to, we use the cosmological constraints that are obtained from lensing data in a similar setup to what we employ in this paper. For the DES-Y1 \cite{2018PhRvD..98d3526A} and KiDS-1000 \citep{2021A&A...646A.140H} results, we use cosmological constraints from a combination of galaxy-galaxy lensing (``GGL''), galaxy clustering (``GGC'') and cosmic shear (``CS''). These lensing surveys cover a wider area than does the HSC-Y1 ($\sim 1000$~deg$^2$ compared to $140$~deg$^2$), but are shallower than is HSC-Y1. We used public results, available from the websites \footnote{\url{http://desdr-server.ncsa.illinois.edu/despublic/y1a1_files/chains/wg_l3.txt}} and \footnote{\url{http://kids.strw.leidenuniv.nl/DR4/data_files/KiDS1000_3x2pt_fiducial_chains.tar.gz}}, for DES-Y1 and KiDS-1000, respectively. We note that the DES-Y1 result was obtained by varying the neutrino mass. However, this difference is not important because the neutrino mass is not well constrained by lensing information or large-scale structure information alone. We refer to the ``CS''$\times$``GGL'' result for KiDS-1000, as that KiDS-1000 analysis used redshift-space galaxy clustering information, and included the BAO information. Hence, Fig.~\ref{fig:corner-comparion-externals} tries to give an apple-to-apple comparison between the lensing results. 

Our cosmological results are weaker than the other constraints we show, because our method only uses large-scale clustering information in the quasi-nonlinear regime where the minimal bias model is valid. In our companion paper \citep{Miyatake:2021b}, we incorporate smaller-scale data using an HOD model.  As shown in Fig.~\ref{fig:corner-comparion-externals}, the resulting constraints, denoted as ``HSC-Y1 GGC$\times$GGL, Miyatake HOD'', become comparable with the results of other lensing experiments, despite the fact that the solid angle covered by HSC-Y1 is 10 times smaller than that of DES-Y1 or KiDS-1000. Hence our constraints can be considered as a conservative result. In addition, our result is consistent with the results of both {\it Planck} 2018 and the HSC-Y1 cosmic shear analysis.

\begin{figure}
\begin{center}
    \includegraphics[width=\columnwidth]{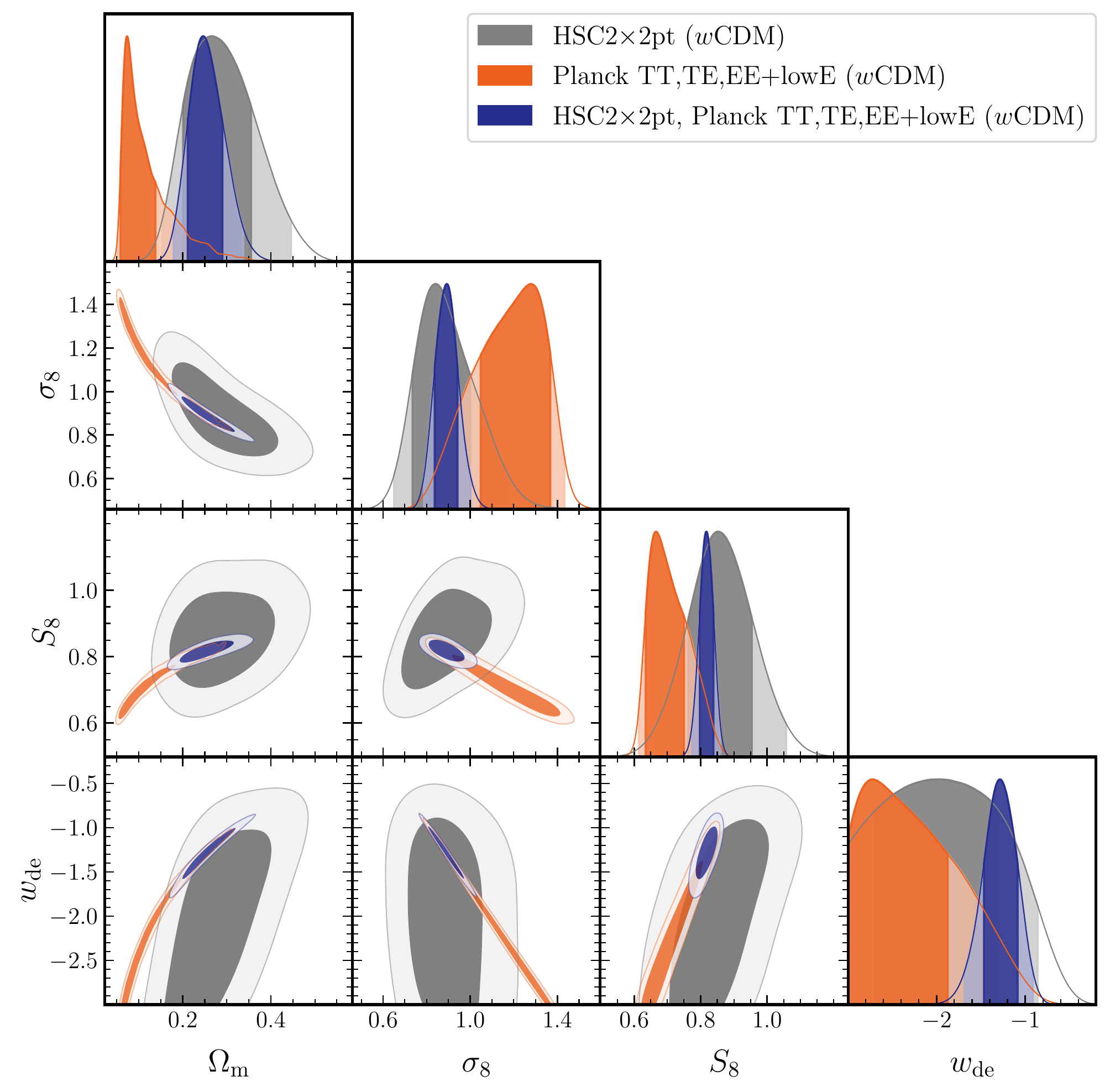}
\end{center}
\caption{The posterior distributions of parameters including the dark energy of equation parameter, $w_{\rm de}$,  for the joint parameter inference of the HSC-Y1, SDSS and {\it Planck2018} data for flat-geometry $w$CDM cosmologies. For {\it Planck}, we include the primary anisotropy data for temperature and $E$-mode polarization, denoted as ``TT,TE,EE+lowE'' \citep[see][]{Aghanim:2018eyx}. That is, we did not include the CMB lensing information.}
\label{fig:corner-wcdm}
\end{figure}

\begin{figure}
\begin{center}
    \includegraphics[width=\columnwidth]{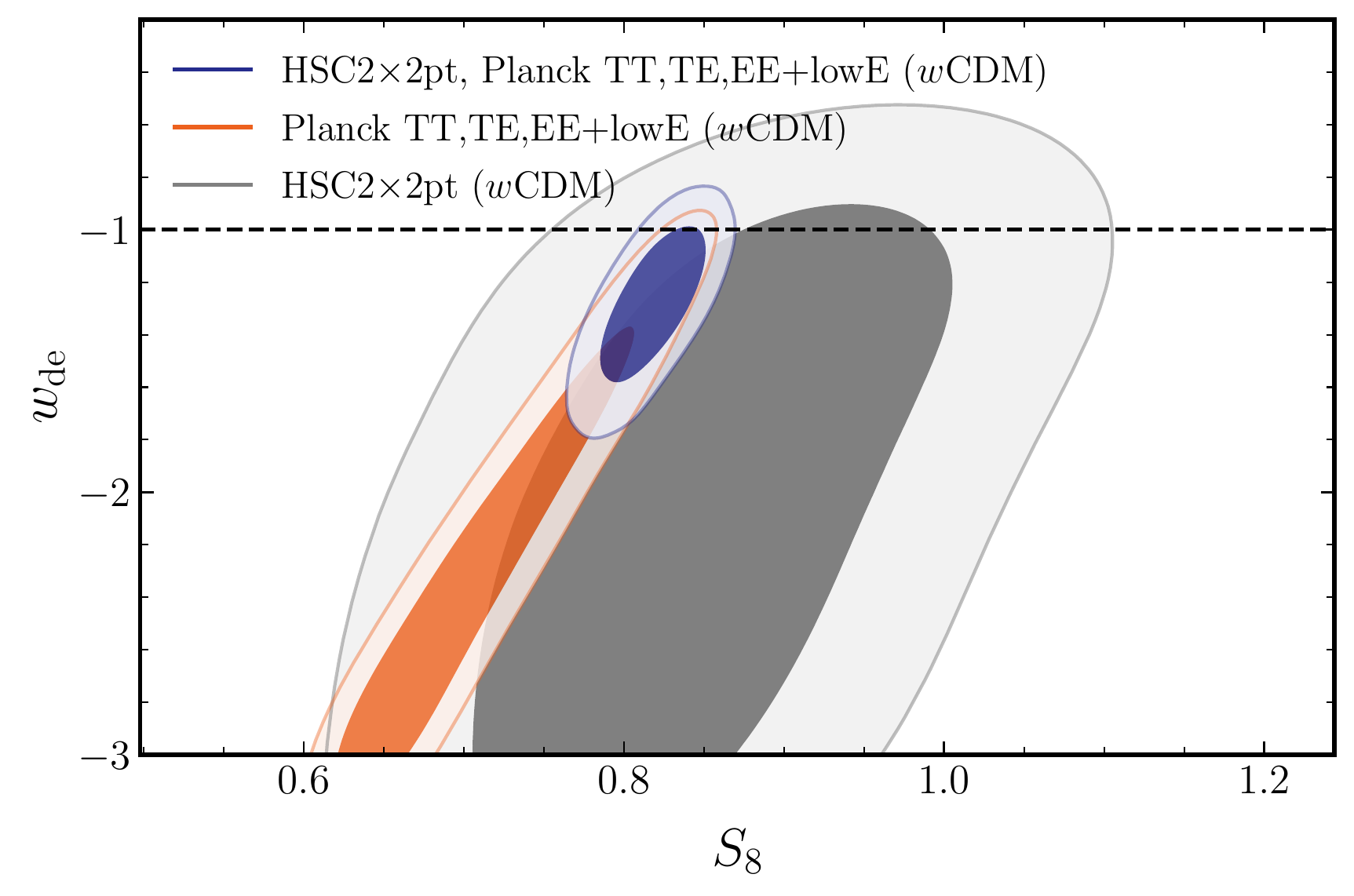}
\end{center}
\caption{Similar to the previous figure, now showing a zoom-in version of the posterior distribution in the sub-space of $(w_{\rm de}, S_8)$.}
\label{fig:wcdm-wde-s8}
\end{figure}

In Fig.~\ref{fig:corner-wcdm}, we show the cosmological parameters estimated from the joint analysis of our HSC-Y1 likelihood and the {\it Planck} 2018 likelihood for a flat $w$CDM model. Here we add the dark energy equation of state parameter, $w_{\rm de}$, in addition to parameters we have so far used for the $\Lambda$CDM model. We employ the priors described in Section~\ref{sec:bayes} and Table~\ref{tab:parameters} except for $\omega_{\rm b}$ and $n_{\rm s}$, for which we employ uniform priors ${\cal U}(0.02190, 0.02285)$ and ${\cal U}(0.9500,0.9781)$, respectively. We use the public {\it Planck} 2018 likelihood code, for the primary anisotropy information (``TT,EE,TE+lowE''), to perform the joint analysis. Since the {\it Planck} information strongly constrains some of the cosmological parameters such as $\omega_{\rm b}$, $\omega_{\rm c}$ and $n_s$, the joint analysis helps breaking parameter degeneracies. Nevertheless, we note that the {\it Planck} information alone cannot constrain $w_{\rm de}$: the posterior distribution of $w_{\rm de}$ for {\it Planck} alone extends to the lower edge of the prior range. We also note that the addition of $w_{\rm de}$ causes strong degeneracies between the cosmological parameters, reflecting that the {\it Planck} information alone cannot constrain these parameters simultaneously. Hence we should not consider seriously about the consistency between the {\it Planck} result and our result. On the other hand, when the HSC constraints in the local universe is combined with the {\it Planck} constraints, it allows us to infer the growth of large-scale structure and then use it to constrain the equation of state parameter of dark energy, $w_{\rm de}$. The joint analysis now shows tightened constraints on $S_8$, $\Omega_{\rm m}$ and $\sigma_8$ (see Table~\ref{tab:summary-table}). The joint analysis prefers a value of $w_{\rm de}$ slightly smaller than $-1$, but the deviation from this value is not significant (also see Fig.~\ref{fig:wcdm-wde-s8}), meaning that the inferred model is consistent with a flat $\Lambda$CDM model.

\section{Summary}
\label{sec:summary-and-conclusion}

In this paper we have presented cosmological constraints from a joint-probe cosmology analysis combining the galaxy-galaxy weak lensing and the projected correlation function, measured from the HSC-Y1 imaging galaxy catalog and the spectroscopic SDSS galaxy catalog. To do this, we adopted a conservative standing point: we employed a perturbation theory based model, more specifically the ``minimal bias'' model, as the theoretical template to interpret the clustering observables. As shown in our validation paper \citep{Sugiyama:2020kfr}, this method can properly extract the cosmological parameters in an unbiased way {\it as long as} the analysis is restricted to large scales ($R\gtrsim 10\,h^{-1}{\rm Mpc}$), as nonlinear effects such as nonlinear clustering and baryonic physics are confined to local scales. This method is also robust against the complication of assembly bias, because assembly bias changes the clustering amplitudes on large scales in such a way that the cross-correlation coefficient function $r_{\rm gg}(r) \equiv\xi_{\rm gm}(r)/[\xi_{\rm gg}(r)\xi_{\rm mm}(r)]^{1/2}$ is close to unity \citep{2021MNRAS.501.1603H}, which is satisfied by the minimal bias model by definition.

For our baseline analysis we employed the BBN prior on $\omega_{\rm b}$ and the {\it Planck} prior on $n_{\rm s}$, but adopted broad priors on other parameters including the linear galaxy bias parameters, $b_1$, for the three galaxy samples. The cosmological constraints we obtained are summarized as $S_8=0.936^{+0.092}_{-0.086}$, $\sigma_8=0.85^{+0.16}_{-0.11}$, and $\Omega_{\rm m}=0.283^{+0.12}_{-0.035}$ for the flat $\Lambda$CDM model. Thus, encouragingly, the joint-probe cosmology helps lift degeneracies between the cosmological parameters and the bias parameters; the  bias parameters are constrained to a fractional precision of about 20\% for each galaxy sample. However, we found a non-negligible degeneracy between $\omega_{\rm c}$ and the aforementioned parameters. We further combined the HSC-Y1 and SDSS constraints with the {\it Planck} likelihood, and found $S_8=0.817^{+0.022}_{-0.021}$, $\sigma_8=0.892^{+0.051}_{-0.056}$, $\Omega_{\rm m}=0.246^{+0.045}_{-0.035}$ and $w_{\rm de}=-1.28^{+0.20}_{-0.19}$ for the flat $w$CDM model. These parameters are consistent with the flat $\Lambda$CDM model inferred from the {\it Planck} 2018 experiment. The statistical errors on the parameters from our analysis are still significantly larger than those which include small-scale information (\citealt{Miyatake:2021b}). However, we believe that the true cosmological model should be contained within the credible regions of our favored models, if the $\Lambda$CDM framework is correct. In our analysis we found no evidence for residual systematic errors comparable to the current statistical errors.

We can expect further improvements in the cosmological constraints with the upcoming HSC data. The galaxy shape catalog of the HSC Year 3 data, covering 450~\sqdeg\ (3 times that of HSC-Y1) and  described in Ref.~\cite{2021arXiv210700136L}, should considerably improve the statistical precision of the weak lensing measurements. We have now developed the cosmology analysis pipeline, and we will use it on a HSC Year 3 analysis. We will carry out a 3$\times$2pt analysis \citep{2018PhRvD..98d3526A}, combining the HSC and SDSS joint-probe analysis with the HSC cosmic shear information.

\acknowledgments
We would like to thank Rachel~Mandelbaum for very useful comments during this work was done. 
This work was supported in part by World Premier International Research Center Initiative (WPI Initiative), MEXT, Japan, and JSPS KAKENHI Grant Numbers JP15H03654, JP15H05887, JP15H05893, JP15H05896, JP15K21733, JP17H01131, JP17K14273, JP18H04350, JP18H04358, JP19H00677, JP19K14767, JP20H00181, JP20H01932, JP20H04723, 
JP20H05850, JP20H05855, JP20H05861, JP21J00011, JP20H05856, JP21H01081 and JP21J10314 by Japan Science and Technology Agency (JST) CREST JPMHCR1414, by JST AIP Acceleration Research Grant Number JP20317829, Japan, and by Basic Research Grant (Super AI) of Institute for AI and Beyond of the University of Tokyo. SS is supported by International Graduate Program for Excellence in Earth-Space Science (IGPEES), World-leading Innovative Graduate Study (WINGS) Program, the University of Tokyo. HM and MSi were supported by the Jet Propulsion Laboratory, California Institute of Technology, under a contract with the National Aeronautics and Space Administration. KO is supported by JSPS Research Fellowships for Young Scientists. YK was supported by the Advanced Leading Graduate Course for Photon Science at the University of Tokyo.

The Hyper Suprime-Cam (HSC) collaboration includes the astronomical
communities of Japan and Taiwan, and Princeton University. The HSC
instrumentation and software were developed by the National Astronomical
Observatory of Japan (NAOJ), the Kavli Institute for the Physics and
Mathematics of the Universe (Kavli IPMU), the University of Tokyo, the
High Energy Accelerator Research Organization (KEK), the Academia Sinica
Institute for Astronomy and Astrophysics in Taiwan (ASIAA), and
Princeton University. Funding was contributed by the FIRST program from
Japanese Cabinet Office, the Ministry of Education, Culture, Sports,
Science and Technology (MEXT), the Japan Society for the Promotion of
Science (JSPS), Japan Science and Technology Agency (JST), the Toray
Science Foundation, NAOJ, Kavli IPMU, KEK, ASIAA, and Princeton
University.
This paper makes use of software developed for the Vera C. Rubin LSST. 
We thank the LSST Project for making their code available as
free software at \url{http://dm.lsst.org}.

The Pan-STARRS1 Surveys (PS1) have been made possible through
contributions of the Institute for Astronomy, the University of Hawaii,
the Pan-STARRS Project Office, the Max-Planck Society and its
participating institutes, the Max Planck Institute for Astronomy,
Heidelberg and the Max Planck Institute for Extraterrestrial Physics,
Garching, The Johns Hopkins University, Durham University, the
University of Edinburgh, Queen's University Belfast, the
Harvard-Smithsonian Center for Astrophysics, the Las Cumbres Observatory
Global Telescope Network Incorporated, the National Central University
of Taiwan, the Space Telescope Science Institute, the National
Aeronautics and Space Administration under Grant No. NNX08AR22G issued
through the Planetary Science Division of the NASA Science Mission.

\bibliography{./refs}

\appendix
\section{Magnification bias effect}
\label{sec:magbias}
In this appendix, we derive the additional contribution due to magnification bias to the covariance matrix of the lensing profile $\dSigma$. 

The lensing magnification due to large-scale structure between us and lens galaxies causes modulations in the number densities of lens galaxies. As a result, the observed number density fluctuation field of lens galaxies is given by
\begin{align}
    \delta_{\rm g}(\chi_{\rm l}, \chi_{\rm l}\bm{\theta}) 
    &= \delta_{\rm g}^{\rm int}(\chi_{\rm l}, \chi_{\rm l}\bm{\theta}) \nonumber\\
    &\hspace{2em}+2(\alpha_{\rm mag, l}-1)\kappa(\chi_{\rm l}, \chi_{\rm l}\bm{\theta})
\end{align}
where $\delta_{\rm g}^{\rm int}$ is the intrinsic number density fluctuation field, $\alpha_{\rm mag}$ is the power-law slope of galaxy counts around a given magnitude cut (see main text), and $\kappa(\chi_{\rm l}\bm{\theta}, \chi_{\rm l})$ is the convergence field that is the projected mass density field up to $z_{\rm l}$ in the direction $\bm{\theta}$, given by
\begin{align}
    \kappa(\chi_{\rm l}, \chi_{\rm l}\bm{\theta}) = \int_0^{\chi_{\rm l}}{\rm d}\chi W(\chi, \chi_{\rm l}) \delta(\chi, \chi_{\rm l}\bm{\theta})
\end{align}
with the lensing efficiency function 
\begin{align}
W(\chi, \chi_{\rm l})\equiv \frac{3}{2}H_0^2\Omega_{\rm m}a^{-1}\chi \frac{\chi_{\rm l}-\chi}{\chi_{\rm l}}.
\end{align}
The same large-scale structure, characterized by $\kappa$, also distorts images of source galaxies used in the galaxy-galaxy weak lensing measurements, and therefore the magnification bias adds additional statistical scatter to the measurements. 

Extending the method in \citet{Oguri:2010vi} and \citet{2018MNRAS.478.4277S}, we derive the additional contribution due to magnification bias to the covariance matrix between the lensing profiles for the lens samples at $z_{\rm l}$ and $z_{\rm l'}$,  i.e. $\dSigma(R_n; z_{\rm l})$ and $\dSigma(R_{n'}; z_{\rm l'})$: 
\begin{widetext}
\begin{align}
    &\hspace{-2em}\delta {\rm Cov}[\dSigma(R_{n},z_{\rm l}),\dSigma(R_{n'},z_{\rm l'})]\nonumber\\
    =&\frac{1}{\Omega_{\rm s}}\int\!\frac{\ell\mathrm{d}\ell}{2\pi}~
    \hat{J}_2\!\left(\ell\frac{R_n}{\chi_{\rm l}}\right)\hat{J}_2\!\left(\ell\frac{R_{n'}}{\chi_{\rm l'}}\right)\Sigma_{\rm cr}(z_{\rm s}, z_{\rm l})\Sigma_{\rm cr}(z_{\rm s}, z_{\rm l'}) \nonumber\\
    &\times\left\{
    \left(2(\alpha_{\rm l'}-1)C_{{\rm g}\kappa_{\rm l}}(\ell; z_{\rm l}, z_{\rm l'})+2(\alpha_{\rm l}-1)C_{{\rm g}\kappa_{\rm l}}(\ell; z_{\rm l'},z_{\rm l})\right)
    \left(C_{\kappa_{\rm s}\kappa_{\rm s}}(\ell)+\frac{\sigma_\epsilon^2}{\bar{n}_{\rm s}}\right)\right. \nonumber\\
    &\hspace{2em}+
    2(\alpha_{l'}-1)C_{{\rm g}\kappa_{\rm s}}(\ell; z_{\rm l})C_{\kappa_{\rm l}\kappa_{\rm s}}(\ell; z_{\rm l'}) + 2(\alpha_{\rm l}-1)C_{\kappa_{\rm l}\kappa_{\rm l}}(\ell; z_{\rm l})C_{{\rm g}\kappa_s}(\ell; z_{\rm l'})\nonumber\\
    &\left.\hspace{2em}+4(\alpha_{\rm l}-1)(\alpha_{\rm l'}-1)
    \left[
    C_{\kappa_{\rm l}\kappa_{\rm l}}(\ell; z_{\rm l},z_{\rm l'})\left(C_{\kappa_{\rm s}\kappa_{\rm s}}(\ell)+\frac{\sigma_\epsilon^2}{\bar{n}_{\rm s}}\right)
    +C_{\kappa_{\rm l}\kappa_{\rm s}}(\ell; z_{\rm l})C_{\kappa_{\rm l}\kappa_{\rm s}}(\ell; z_{\rm l'})
    \right]\right\}.
    \label{eq:cov_magnificationbias}
\end{align}
In this derivation, we assumed that the lens galaxies are at a single redshift, $z_{\rm l}$ or $z_{\rm l'}$, for simplicity, and we consider the case with $z_{\rm l'}\geq z_{\rm l}$ without loss of generality. The auto- and cross-angular power spectra for the lensing convergence field and the projected field of lens galaxies in the above equation are defined as
\begin{align}
    &C_{{\rm g}\kappa_{\rm s}}(\ell; z_{\rm l}) = \frac{W(\chi_{\rm l}, \chi_{\rm s})}{\chi_{\rm l}^2}P_{\rm gm}\left(\frac{\ell}{\chi_{\rm l}}, z_{\rm l}\right)\\
    &C_{{\rm g}\kappa_{\rm l}}(\ell; z_{\rm l}, z_{\rm l'}) = \frac{W(\chi_{\rm l},\chi_{\rm l'})}{\chi_{\rm l}^2} P_{\rm gm}\left(\frac{\ell}{\chi_{\rm l}}, z_{\rm l}\right)\Theta(z_{\rm l'}-z_{\rm l}) \label{eq:c_gzldashed}\\
    &C_{\kappa_{\rm l}\kappa_{\rm s}}(\ell, z_{\rm l}) = \int{\rm d}\chi \frac{W(\chi, \chi_{\rm s})W(\chi, \chi_{\rm l})}{\chi^2}P_{\rm mm}^{\rm NL}\left(\frac{\ell}{\chi},z\right)\\
    &C_{\kappa_{\rm s}\kappa_{\rm s}}(\ell) = \int{\rm d}\chi\frac{W(\chi, \chi_{\rm s})^2}{\chi^2}P_{\rm mm}^{\rm NL}\left(\frac{\ell}{\chi}, z_{\rm l}\right)\\
    &C_{\kappa_{\rm l}\kappa_{\rm l}}(\ell, z_{\rm l}, z_{\rm l'}) = \int{\rm d}\chi\frac{W(\chi, \chi_{\rm l})W(\chi, \chi_{\rm l'})}{\chi^2}P_{\rm gm}^{\rm NL}\left(\frac{\ell}{\chi}, z\right),
\end{align}
\end{widetext}
where $\Theta(x)$ is the Heaviside step function; $\Theta(x)=1$ if $x>0$, otherwise $\Theta(x)=0$, and $\hat{J}_2(x)$ is the 2nd-order Bessel function averaged within the separation bin, defined for the $n$-th separation bin ($R_n$) as
\begin{align}
    &\hat{J}_2(kR_n) = \frac{2}{R_{\rm n,min}^2-R_{\rm n,max}^2}\int_{R_{\rm n,min}}^{R_{\rm n,max}}R{\rm d}R~J_2(kR).
\end{align}
with $k=\ell/\chi$ in Eq.~(\ref{eq:cov_magnificationbias}). Note that we employed Limber's approximation \citep{1953ApJ...117..134L} for the above angular power spectra. We also note that the cross-correlation given by $C_{{\rm g}\kappa_{\rm l'}}(\ell; z_{\rm l},z_{\rm l'})$ (Eq.~\ref{eq:c_gzldashed}) arises because the convergence field, which causes the density modulations for the lens sample at $z_{\rm l'}$ ($\geq z_{\rm l}$), includes the contribution of the mass distribution at $z_{\rm l}$, leading to a cross-correlation between the mass distribution and the lens galaxy distribution that are both at $z_{\rm l}$. The additional contribution due to the lensing magnification turns out not to be significant compared to the total power of the covariance matrix, but we included it in our cosmology analysis for completeness.

\section{Convergence tests of nested sampling}
\label{sec:nestcheck}
We use \texttt{MultiNest} for sampling in our parameter estimation. There are uncertainties in the parameter estimation due to the sampling process itself, because the size of the sampled chain is finite and the sampling itself depends on the seed of \texttt{MultiNest}. In order to make sure that the uncertainty due to \texttt{MultiNest}, $\Delta^{\rm MN}$, is small compared to the statistical uncertainty, $\sigma^{\rm stat.}$, we run eight independent chains with different \texttt{MultiNest} seeds. We confirm that the ratio of the sampler uncertainty to the statistical error is $\Delta^{\rm MN}_{S_8}/\sigma^{\rm stat.}_{S_8}=8.3\times10^{-5}$; thus this source of uncertainty is negligible. 

\begin{figure}
\begin{center}
    \includegraphics[width=\columnwidth]{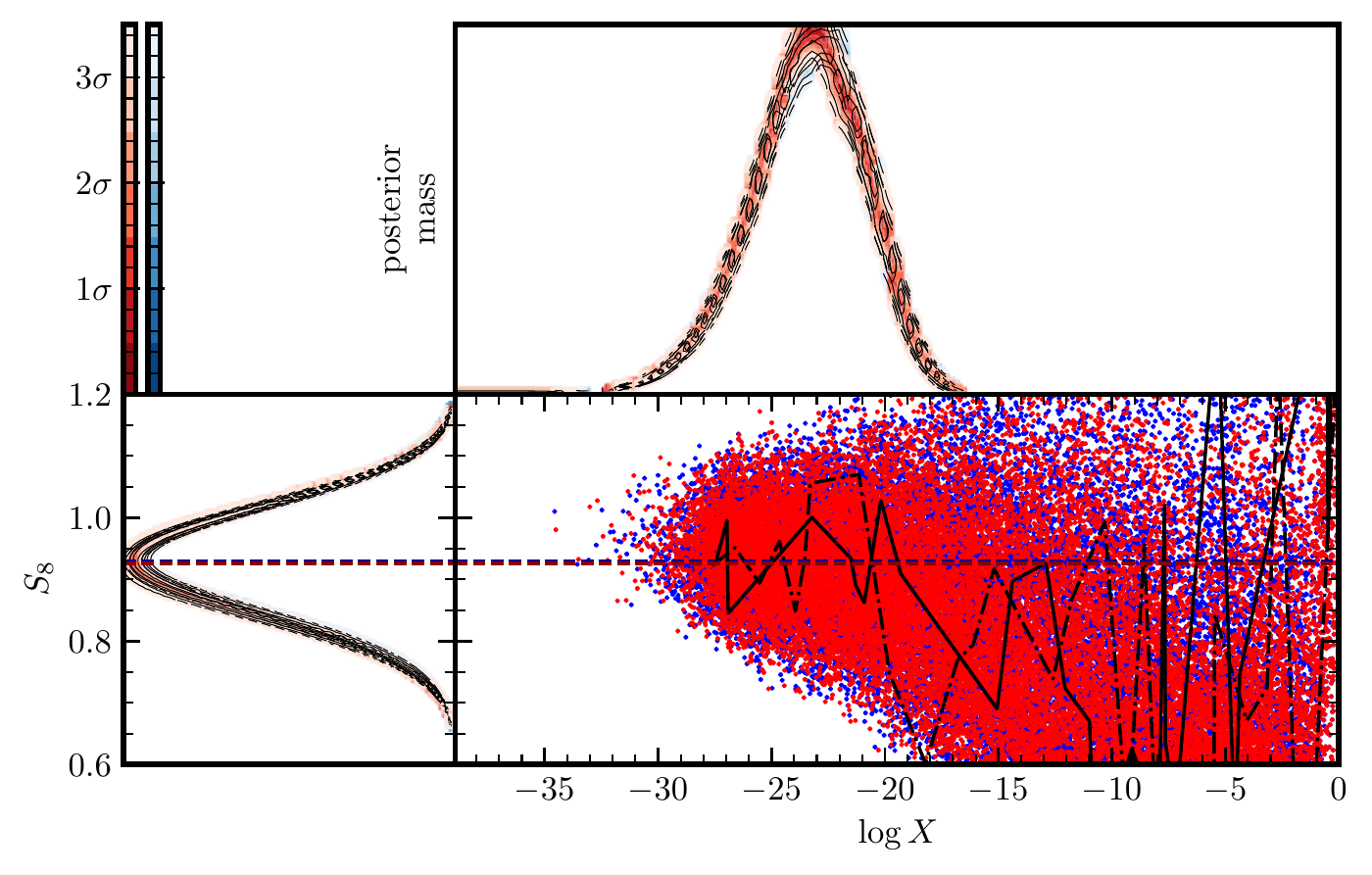}
\end{center}
\caption{The result of \texttt{nestcheck} \citep{2019MNRAS.483.2044H}. Two independent chains with different seeds indicated with different colors are used. The upper right panel shows the posterior mass as a function of logarithmic prior volume corresponding to the iteration number of \texttt{MultiNest}, where the prior volume, $\log X$, decreases as the iteration number increases. The contour shows the uncertainty of the posterior mass of the chain estimated from the bootstrap from each chain. The lower left panel shows the posterior distribution of the $S_8$ value, with uncertainty contour estimated from bootstrap resamples of the chain. The lower right panel shows the 2-d distribution of $\log X$ and $S_8$.}
\label{fig:nestcheck}
\end{figure}

We also check that our choice of the \texttt{MultiNest} hyper-parameters are suitable for sampling the target posterior distribution by using the public code, \texttt{nestcheck} \citep{2019MNRAS.483.2044H}. The result is shown in Fig.~\ref{fig:nestcheck}, which shows the results of two independent chains. We can see that both chains are well converged to the peak of the $S_8$ distribution, there is no significant difference between these chains in the $(\log X, S_8)$ plane, and the uncertainty of the marginalized posterior distribution of $S_8$ estimated from bootstrap resamples of either chain is not large. Thus we conclude that the chains have fully converged with our choice of \texttt{MultiNest} hyper-parameters.

\section{Posterior distributions in a full parameter space}
\label{sec:full-corner-plot}
\begin{figure*}
\begin{center}
    \includegraphics[width=2\columnwidth]{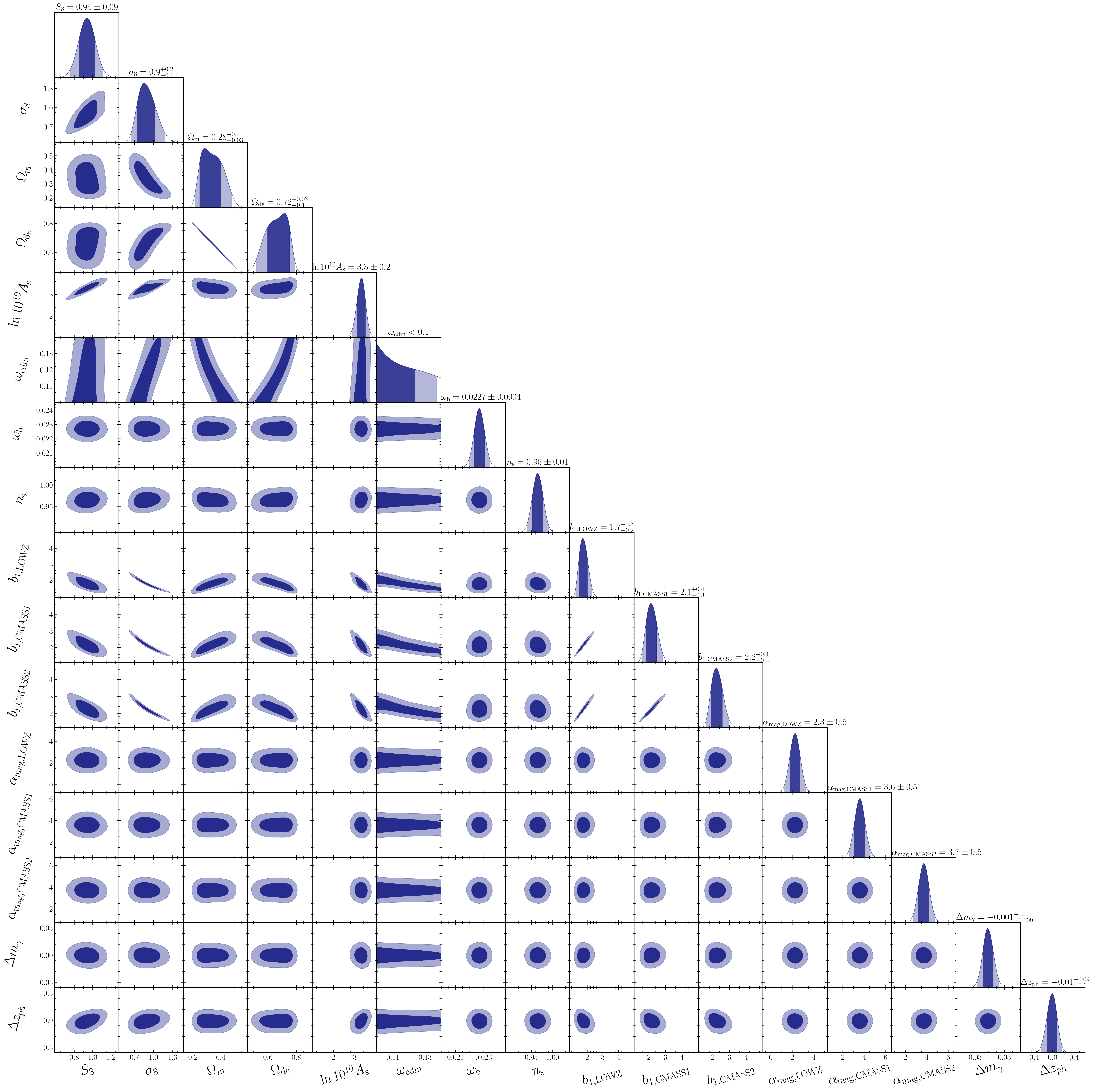}
\end{center}
\caption{The 1-d and 2-d posterior distributions over the full parameter space for the baseline analysis shown in Fig.~\ref{fig:corner-baseline}.}
\label{fig:corner-baseline-full}
\end{figure*}

Fig.~\ref{fig:corner-baseline-full} shows the 1-d and 2-d posterior distributions in a full parameter space in our baseline analysis, including the derived parameters, $\Omega_{\rm m}, \sigma_8$ and $S_8$.

\section{Supplementary information for the results of different analysis setups}
\label{sec:additional-tests}
In this appendix, we show supplemental systematic tests to explore some of the systematics we found in Section~\ref{sec:cosmology-constraint}.

\subsection{Flat peak in $P(\Omega_{\rm m})$ of baseline analysis}\label{sec:Om-bimodal}
\begin{figure}
\begin{center}
    \includegraphics[width=\columnwidth]{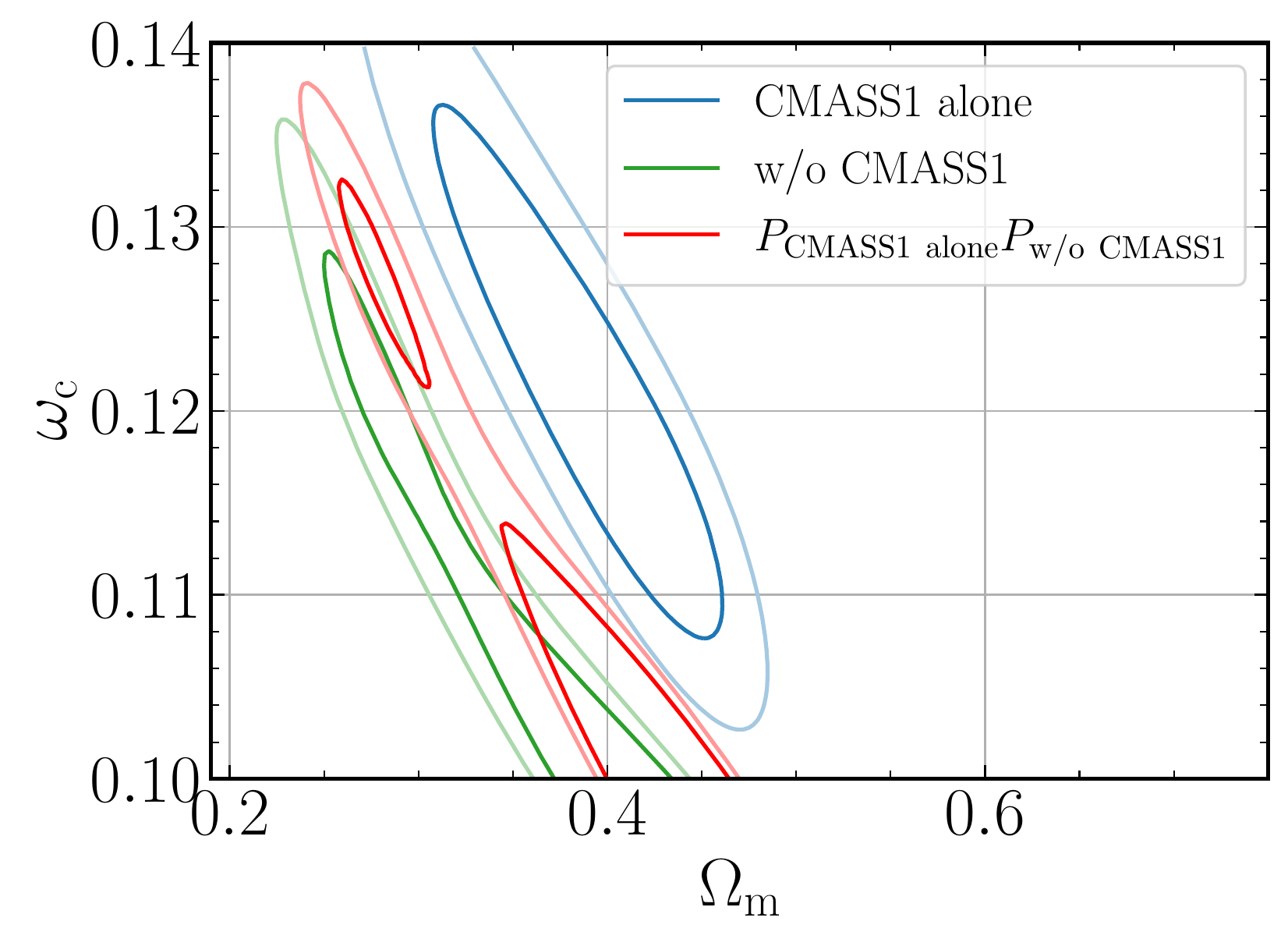}
\end{center}
\caption{Marginalized posterior distributions (20 and 40\% credible regions) of the ``CMASS1 alone'' analysis (blue) and ``w/o CMASS1'' analysis (green) in the $(\Omega_{\rm m},\omega_{\rm c})$ plane. The red contour is the product of the two marginalized posteriors, which shows two peaks in $\Omega_{\rm m}$.}
\label{fig:corner-combine-Omoc}
\end{figure}

For our baseline results shown in Fig.~\ref{fig:corner-baseline}, the 1-d marginalized posterior of $\Omega_{\rm m}$ exhibits a flat-shaped peak. We find that this is caused by cosmic variance between the three lens samples. If we ignore the correlation between the samples, the posterior for the combined samples is equivalent to the product of the posteriors from each sample alone.  This is illustrated in Fig.~\ref{fig:corner-combine-Omoc}, which shows the posterior (20 and 40\% credible regions) in the $\Omega_{\rm m}-\omega_{\rm c}$ subspace for the CMASS1-alone analysis and the analysis without the CMASS1 sample.  The contours of the two do not overlap. The joint posterior has peaks at low and high values of $\Omega_{\rm m}$, which explains why we find a flat-shaped peak when we marginalize over $\omega_{\rm c}$.

\subsection{A test for systematic effects in Ephor AB}\label{sec:ephor-test}
\begin{figure}
\begin{center}
    \includegraphics[width=\columnwidth]{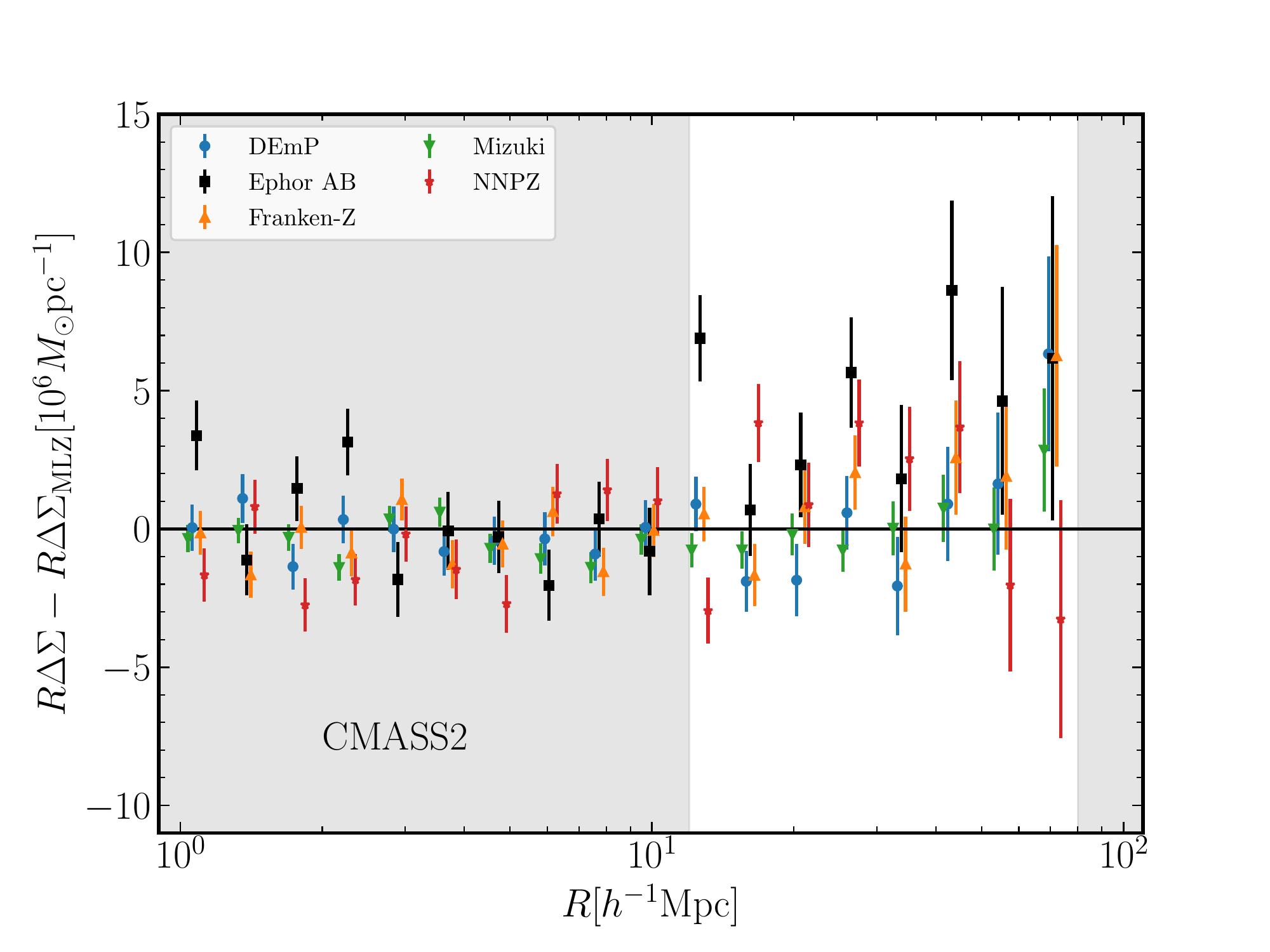}
\end{center}
\caption{The difference between the galaxy-galaxy weak lensing signals of the CMASS2 sample using different photo-z catalogs. Because the number of source-lens pairs in the Ephor AB catalog is $\sim 35$\% smaller than that of the fiducial (MLZ) catalog, the signal in Ephor AB is noisy, and happens to be a several-sigma deviation from that in the fiducial catalog.}
\label{fig:ephor-dSigma-diff}
\end{figure}

\begin{figure}
\begin{center}
    \includegraphics[width=\columnwidth]{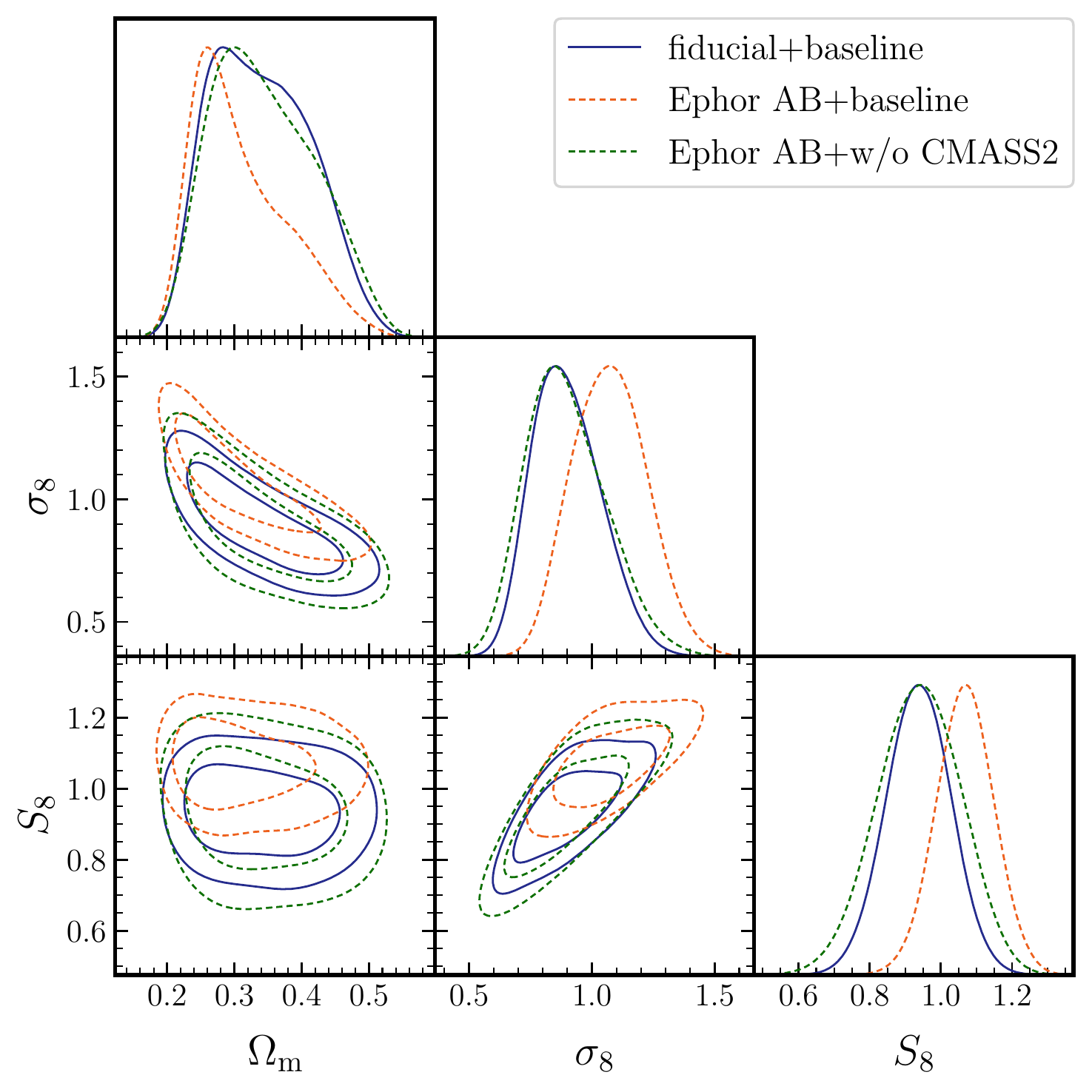}
\end{center}
\caption{Comparison of cosmological parameter constraints from the signal measured in the Ephor AB catalog to that of the fiducial catalog. When the noisy CMASS2 sample in the Ephor AB catalog is removed from the data vector, the constraints on $\sigma_8$ and $S_8$ are consistent with the fiducial result.}
\label{fig:corner-ephor}
\end{figure}

As shown in Fig.~\ref{fig:bar-plot}, when we used the Ephor~AB photo-$z$ catalog we found significantly larger values of $\sigma_8$ and $S_8$ than in our baseline analysis. We found that the number of source-lens pairs for the CMASS2 sample in the Ephor~AB catalog is $\sim 35\%$ smaller than that in the fiducial photo-$z$ catalog (MLZ), and the lensing signal measured in Ephor~AB catalog has larger statistical errors. In particular, Fig.~\ref{fig:ephor-dSigma-diff} shows that the galaxy-galaxy lensing signal for the CMASS2 sample, measured from the Ephor AB catalog, displays larger amplitudes at $R>12[h^{-1}{\rm Mpc}]$ than do the other catalogs, which explains why we find larger values of $\sigma_8$ and $S_8$. For a further check we confirm that the constraints on $\sigma_8$ and $S_8$ for the Ephor~AB catalog become consistent with the results of our baseline analysis if we do not use the lensing signal of the CMASS2 sample in the parameter inference, as shown in Fig.~\ref{fig:corner-ephor}.

\end{document}